\documentclass[journal]{IEEEtran}

\usepackage{amsfonts}
\usepackage{array}
\usepackage[caption=false,font=normalsize,labelfont=sf,textfont=sf]{subfig}
\usepackage{stfloats}
\usepackage{verbatim}
\usepackage{graphicx}
\usepackage{cite}
\usepackage[table,xcdraw]{xcolor}
\usepackage{textcomp}
\usepackage{url}
\usepackage{microtype}
\usepackage{algorithmicx}
\usepackage[linesnumbered, ruled, boxed]{algorithm2e}
\usepackage{epstopdf}
\usepackage{listings}
\usepackage{mdframed}
\usepackage{makecell,multirow}
\usepackage{url}
\usepackage{multicol}

\usepackage{float}
\usepackage{balance}
\usepackage{pifont}
\usepackage{enumitem}
\usepackage{adjustbox}
\usepackage{afterpage}

\usepackage[normalem]{ulem}
\usepackage{tcolorbox}
\usepackage{threeparttable}
\usepackage{tcolorbox}
\usepackage{amsmath}
\usepackage{booktabs}

\newcommand{\code}[1]{\lstinline|#1|}

\newcommand{\tech}{KUMIC}

\newcounter{mycounter}
\newcounter{finding}

\newcommand{\distance}{2pt}
\definecolor{top1}{gray}{0.55}
\definecolor{top2}{gray}{0.65}
\definecolor{top3}{gray}{0.75}
\definecolor{top4}{gray}{0.85}
\definecolor{top5}{gray}{0.9}

\newcounter{commentcounter} 

\newcounter{modifytext}
\colorlet{colorchangeframe}{black!20}
\colorlet{colorchangebg}{black!2}

\begin{document}
\title{Optimizing Knowledge Utilization for Multi-Intent Comment Generation with Large Language Models}

\author{Shuochuan Li, Zan Wang, Xiaoning Du, Zhuo Wu, Jiuqiao Yu, Junjie Chen

\thanks{Shuochuan Li, Zan Wang,  Zhuo Wu  are with School of Computer Software and School of New Media and Communication, Tianjin University, Tianjin, China. E-mail:\{lishuochuan, wangzan, wuzhuo\}@tju.edu.cn. }

\thanks{Junjie Chen are with School of Computer Software, Tianjin University, Tianjin, China. E-mail:junjiechen@tju.edu.cn. Junjie Chen is the corresponding author.}

\thanks{Xiaoning Du is with Faculty of Information Technology, Monash University, Australia. E-mail:xiaoning.du@monash.edu.}

\thanks{Jiuqiao Yu is with UC Berkeley Engineering, University of California, Berkeley. E-mail:jiuqiao\_yu@berkeley.edu.}
}


\markboth{IEEE TRANSACTIONS ON SOFTWARE ENGINEERING,~Vol.~14, No.~8, July~2024}%
{Shell \MakeLowercase{\textit{et al.}}: A Sample Article Using IEEEtran.cls for IEEE Journals}


\maketitle

\begin{abstract}
Code comment generation aims to produce a generic overview of a code snippet, helping developers understand and maintain code.
However, generic summaries alone are insufficient to meet the diverse needs of practitioners; for example, developers expect the implementation insights to be presented in an untangled manner, while users seek clear usage instructions.
This highlights the necessity of multi-intent comment generation.
With the widespread adoption of Large Language Models (LLMs) for code-related tasks, these models have been leveraged to tackle the challenge of multi-intent comment generation. 
Despite their successes, state-of-the-art LLM-based approaches often struggle to construct correct relationships among intents, code, and comments within a smaller number of demonstration examples.
To mitigate this issue, we propose a framework named \tech{} for multi-intent comment generation.
Built upon in-context learning, \tech{} leverages Chain-of-Thought (CoT) to optimize knowledge utilization for LLMs to generate intent-specific comments.
Specifically, \tech{} first designs a retrieval mechanism to obtain similar demonstration examples, which exhibit high code-comment consistency.
Then, \tech{} leverages CoT to guide LLMs to focus on statements facilitating the derivation of code comments aligned with specific intents.
In this context, \tech{} constructs a mapping knowledge chain — linking code to intent-specific statements to comments — which enables LLMs to follow similar reasoning steps when generating the desired comments.
We conduct extensive experiments to evaluate \tech{}, and the results demonstrate that \tech{} outperforms state-of-the-art baselines by 14.49\%, 22.41\%, 20.72\%, and 12.94\% in terms of BLEU, METEOR, ROUGE-L, and SBERT, respectively.

\end{abstract}


\section{Introduction}
\label{sec:introduction}

\begin{table*}[t]
\centering
\belowrulesep=0pt
\aboverulesep=0pt
\caption{The intent taxonomy of code comments\label{tab:intents} }
\begin{tabular}{c|l|l}
\toprule
\textbf{Category} & \multicolumn{1}{c|}{\textbf{Definition}} & \multicolumn{1}{c}{\textbf{Example}} \\
    \midrule
    What  & \makecell[l]{Describes the functionality of a method} & \makecell[l]{Starts the background initialization.} \\
    \midrule
    Why   & \makecell[l]{Explains the reason why a method is provided \\or the design rationale of the method} & \makecell[l]{With this method the initializer becomes active and\\ invokes the initialize() method in a background task.} \\
    \midrule
    How-to-use & \makecell[l]{Describes the usage or the expected set-up of\\ using a method} & \makecell[l]{After the construction of a BackgroundInitializer()\\ object, its start() method has to be called.} \\
    \midrule
    How-it-is-done & \makecell[l]{Describes the implementation details of a method} & \makecell[l]{Get an external executor to create a background task.\\ If there is not any, it creates a new one.} \\
    \midrule
    Property & \makecell[l]{Asserts properties of a method including pre-\\conditions or post-conditions of a method} & \makecell[l]{Return a flag whether the initializer could be started\\ successfully.} \\
    \midrule
    Others & \makecell[l]{Unspecified or ambiguous comments} & \makecell[l]{The implementation is awesome.} \\
\bottomrule
\end{tabular}
\end{table*}

Code comments play a crucial role in software development and maintenance by providing brief, natural-language descriptions of code snippets~\cite{DBLP:journals/tse/XiaBLXHL18,DBLP:conf/sigdoc/SouzaAO05} 
They enable developers to comprehend code swiftly and precisely. 
However, in real-world projects, code and comments are often mismatched, with comments frequently being either missing or outdated~\cite{DBLP:conf/kbse/SridharaHMPV10,DBLP:conf/kbse/LiL000J21}.
This issue arises primarily because manually crafting comments is time-consuming.
As a result, the development of automatic code comment generation approaches has become essential. 
Traditional approaches to comment generation typically produce a single, monolithic description of the code. 
However, this often falls short in practice, as developers from different roles and backgrounds seek distinct information from code comments. 
For instance, backend developers focus on the business logic behind a function (i.e., why the function was developed), while frontend developers are more interested in its usage (i.e., how to use the function)~\cite{DBLP:journals/tosem/ChenXHLL21}.
This inefficiency not only slows development but also complicates maintenance tasks.
To address this challenge, there is a growing demand for automatic code comment generation that produces descriptions tailored to specific intents~\cite{DBLP:conf/icse/HuX0WCZ22}.
Such approaches can offer greater control by explicitly organizing comments based on intent, making them more effective and accessible.
This task is known as \textbf{multi-intent comment generation}~\cite{DBLP:conf/icse/GengWD00JML24}.

Recently, initial efforts have been made in this area.
A taxonomy of intents expected in code comments has been established~\cite{DBLP:conf/icse/ZhaiXSTPMXZTZ20, DBLP:journals/tosem/ChenXHLL21, DBLP:conf/icse/MuCSWW23}, covering aspects such as what a function does and how to use it.
Building upon this taxonomy, 
Mu et al.~\cite{DBLP:conf/icse/MuCSWW23} proposed DOME, the first  multi-intent comment generation approach based on a Transformer model~\cite{DBLP:conf/nips/VaswaniSPUJGKP17}.
However, DOME exhibits limited generalizability. 
In contrast, LLMs, pre-trained on vast corpora encompassing diverse developer intents, offer better generalizability and deeper code understanding.
Geng et al.~\cite{DBLP:conf/icse/GengWD00JML24} thus explored using LLMs with in-context learning, retrieving intent-matched code–comment pairs as demonstrations.
However, this approach struggled to consistently outperform DOME unless the number of demonstrations was increased to ten. 
A possible reason is that their retrieval prioritizes similarity between demonstration examples and the given code, but overlooks the degree to which code and comments align in terms of a specific intent. 
Examples that exhibit a clear and explicit correspondence between code and intent are expected to be more beneficial for LLMs in fulfilling the tasks. 
In this work, we define the knowledge embedded in such code-comment pairs as intent-related knowledge, and aim to select and transform examples that demonstrate high-quality intent-related knowledge—namely, a clearer and more explicit alignment between code and intent.
Without such high-quality intent-related knowledge examples, LLMs may be misled by examples that convey ambiguous or irrelevant information, resulting in incorrect comment generation.

To improve the effectiveness of LLM-based multi-intent comment generation, we first address the first problem:
''\textbf{How can we retrieve the demonstration examples that embody high-quality intent-related knowledge?}''.
Developers tend to more easily understand code-comment pairs with strong semantic consistency, as such pairs provide clearer associations between code and comment.
Inspired by this, we combine similarity-based retrieval to ensure semantic relevance with consistency-based filtering to highlight well-aligned code–comment pairs, to ensure the inclusion of intent-related knowledge.
However, such knowledge-embodied examples may still include noise, which can mislead LLMs to focus on irrelevant information during intent-specific comment generation.

Then, we address the second problem: ''\textbf{How can we guide LLMs to focus on intent-related knowledge within examples when generating comments with specific intents?}''.
Different intents are often reflected in particular code statements that act as a bridge between code and comments.
By enriching demonstration examples with intent-specific statements, we construct a mapping knowledge chain — linking code to intent-specific statements to comments — which can better guide LLMs in identifying intent-related knowledge.
To operationalize this process, we incorporate the Chain-of-Thought (CoT)~\cite{DBLP:conf/nips/Wei0SBIXCLZ22}, guiding LLMs to first identify intent-specific statements and then generate corresponding comments step-by-step.

This also raises a third problem: ``\textbf{How can we extract intent-specific important statements from demonstration examples?}''.
Deep learning models are proficient in capturing intricate relationships between inputs and outputs. 
To identify such statements, we explore training tasks that strengthen the association between intent-specific comments and code. 
Among potential options, the code search task serves as a lightweight yet effective solution — it directly models the semantic correspondence between code and intent-specific comments without requiring intensive generative supervision, which we leverage to locate code regions most aligned with the specific intent and extract the corresponding statements.

Overall, we propose \tech{}, a novel LLM-based multi-intent comment generation approach by optimizing knowledge utilization.
It comprises two key components: Example Retriever and Knowledge Augmentation. 
In the Example Retriever, we first retrieve a set of similar examples from the retrieval corpus for a given code snippet. 
We then perform a quality assessment step to evaluate the semantic consistency between the code and its comment for each example, selecting high-quality intent-related knowledge examples as demonstrations.
In the Knowledge Augmentation, we design a code search task to train a lightweight model, leveraging its attention mechanism to extract intent-specific important statements from each demonstration example. 
These key statements are subsequently incorporated into demonstration examples, enabling LLMs to focus on intent-relevant information and generate intent-specific comments with higher accuracy and relevance.

To evaluate \tech{}, we conducted experiments on two large-scale real-world datasets, Funcom~\cite{DBLP:conf/icse/LeClairJM19} and TLC~\cite{DBLP:conf/ijcai/HuLXLLJ18}.
The results show that \tech{} outperforms the two state-of-the-art baselines by 14.49\%, 22.41\%, 20.72\%, and 12.94\% on average across both datasets in terms of BLEU, METEOR, ROUGE-L, and SBERT, respectively.
Notably, \tech{} achieves the highest improvements for the semantically demanding intents (i.e. "\textit{What}" and "\textit{Why}"), while showing moderate gains for the more straightforward intent (“\textit{How-to-use}”).
Additionally, human evaluation and qualitative analysis confirm that the comments generated by \tech{} are more accurate and helpful in assisting developers’ code comprehension..

Our main contributions are outlined as follows:
\vspace{-2pt}
\begin{itemize} 
\item We propose a novel LLM-based approach for multi-intent code comment generation by optimizing knowledge utilization for LLMs. 

\item We introduce an effective example quality assessment mechanism for retrieving knowledge-rich examples, and design a lightweight solution to automatically extract intent-specific important statements within the code for emphasizing intent-related knowledge in examples.

\item We conduct extensive experiments on two large-scale Java datasets, demonstrating that \tech{} significantly outperforms the state-of-the-art approaches.

\item The source code~\cite{our/url} is open-sourced to facilitate our study's replication and its application in extensive contexts.
\end{itemize}

\section{Background and Motivation}

\subsection{Taxonomy of Comment Intent}

Recent research~\cite{DBLP:conf/icse/HuX0WCZ22} indicates that as codebases grow, developers increasingly seek to understand code from multiple dimensions rather than a single perspective.
Multi-intent comments are essential for facilitating maintenance and development by providing insights into various aspects of code.
Therefore, comments should be categorized into finer-grained divisions to better capture underlying intentions.
We adopt the taxonomy proposed by prior studies~\cite{DBLP:journals/tosem/ChenXHLL21, DBLP:conf/icse/MuCSWW23}, which includes six categories, as shown in Table~\ref{tab:intents}.
Following prior work~\cite{DBLP:conf/icse/MuCSWW23, DBLP:conf/icse/GengWD00JML24}, we treat \textit{Others} comments—defined as unspecific or ambiguous—as noise and exclude them from our approach.
Generating multi-intent comments provides developers with a more comprehensive understanding of the code, aiding both development and maintenance.
For example, "\textit{What}" comments explain the function of a method.
"\textit{Why}" comments clarify the design goals behind the method.
"\textit{How-to-use}" comments help developers quickly grasp its usage.
"\textit{How-it-is-done}" comments reveal inconsistencies between comments and implementations.
Lastly, "\textit{Property}" comments examine  assertions and conditions in the code, aiding program verification and testing.

\begin{figure}[t]
  \centering
  \subfloat[High Quality Example\label{fig:high_code_quality}]{%
    \includegraphics[width=0.85\linewidth]{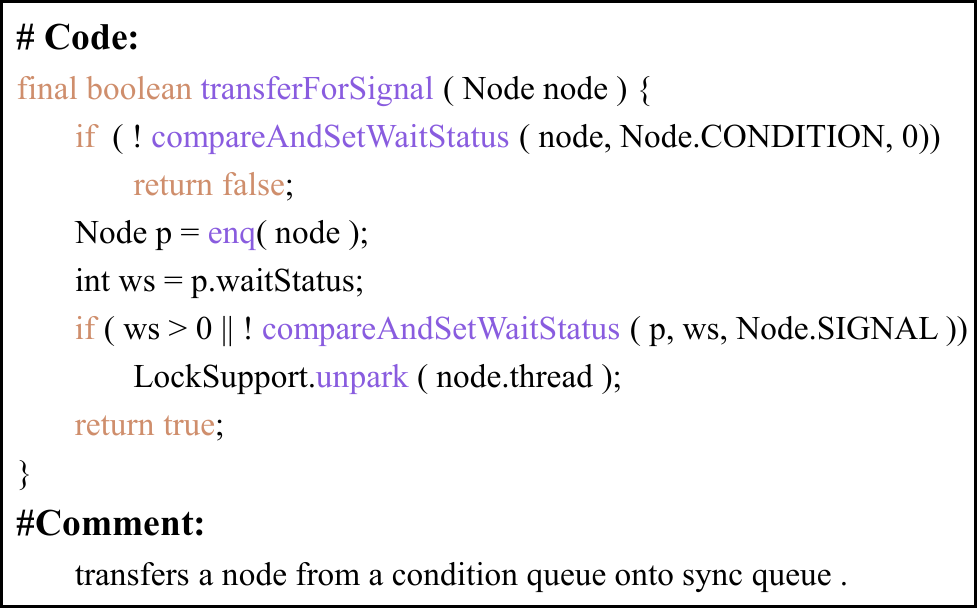}}%
  \\
  \subfloat[Low Quality Example\label{fig:low_code_quality}]{%
    \includegraphics[width=0.85\linewidth]{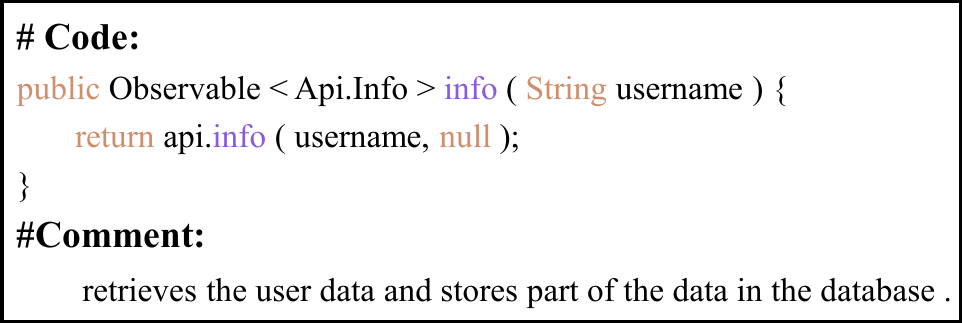}}%
  \caption{Examples of code‑comment pair}
  \label{fig:code_quality_example}
\end{figure}

\subsection{Motivation}

\noindent 
\textbf{Negligence in example quality.}~
Existing in-context learning-based comment generation approaches typically rely on traditional retrieval metrics prioritizing similarity between examples during retrieval. 
However, such metrics neglect example quality, leading to the selection of examples with inconsistent semantics between code and comments. 
For instance, as shown in Fig. \ref{fig:low_code_quality}, the code snippet represents a simple function call, $api.info$, while the associated comment provides rich semantic information.
This discrepancy yields low semantic consistency, as the comment is not grounded in any statements within the code.
Since LLMs must extract task-relevant knowledge from demonstration examples, such low-quality pairs hinder their ability to capture meaningful code–comment correspondences and thus limit intent-specific generation.
To address this limitation, it is crucial to incorporate example quality assessment into the retrieval process by emphasizing semantic consistency between code and comments, ensuring that retrieved examples contain high-quality intent-related knowledge and provide reliable guidance for intent-specific comment generation.

\begin{figure*}[t]
    \centering
    \includegraphics[width=1\linewidth]{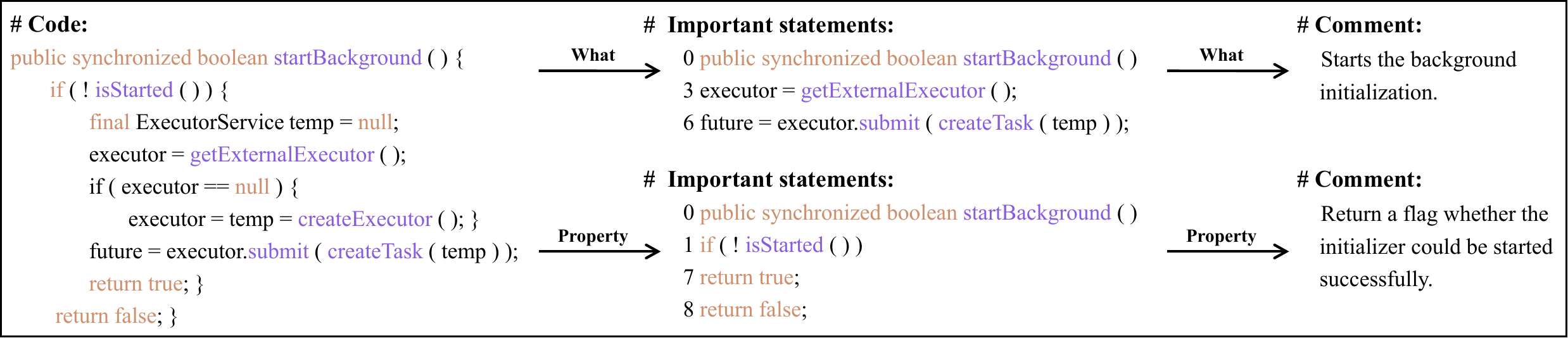}
    \caption{The explicit guidance for LLMs}
    \label{fig:inference}
\end{figure*}

\begin{figure}[t]
  \centering
  \includegraphics[width=1\linewidth]{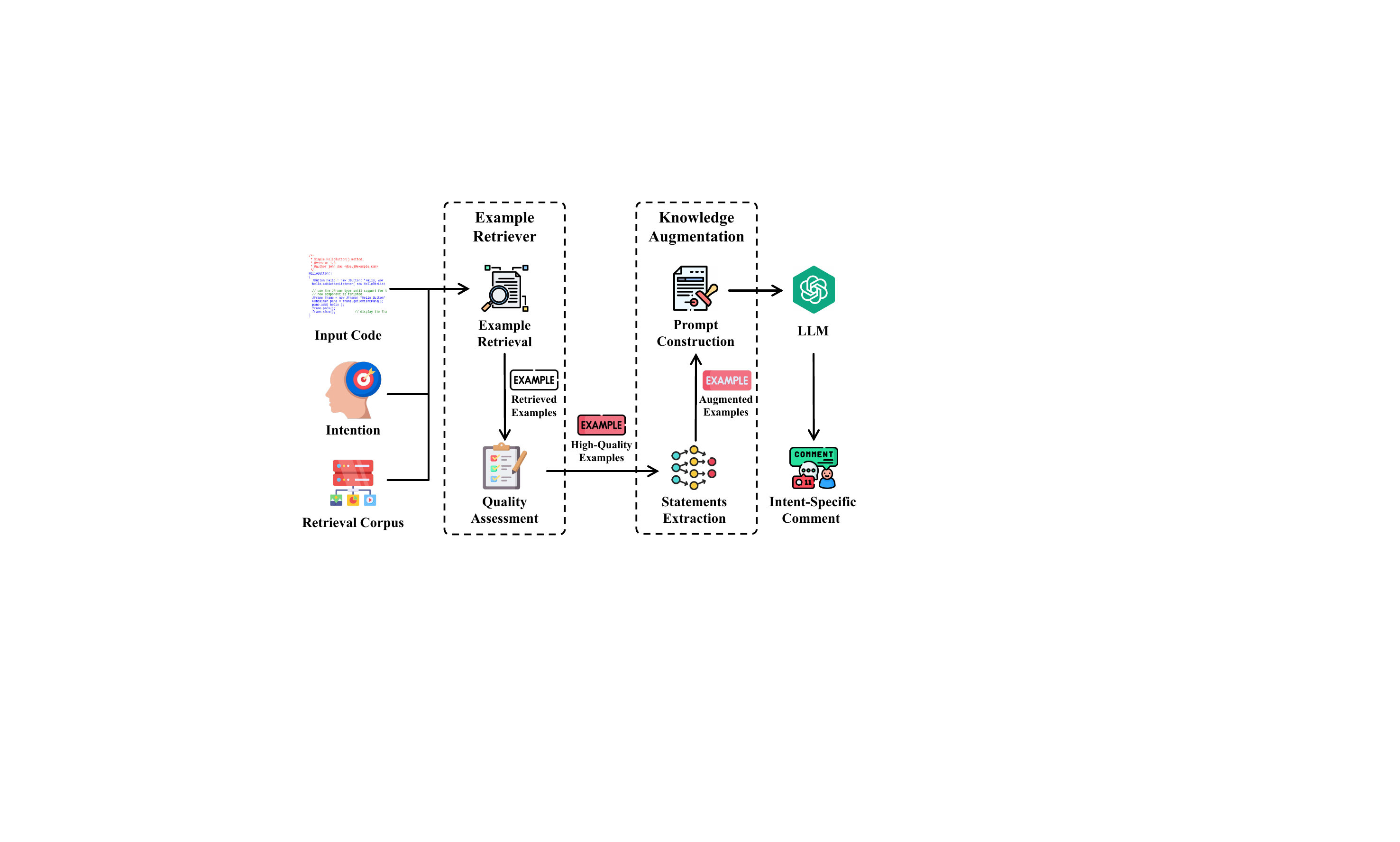}
  \caption{Overview of \tech{}} 
  \label{fig:overview}
\end{figure}

\smallskip 
\noindent 
\textbf{Lack of explicit guidance.}~
Although Geng et al.~\cite{DBLP:conf/icse/GengWD00JML24} mitigate DOME's data scarcity by leveraging LLMs with in-context learning, their approach fails to consistently outperform DOME unless ten demonstration examples are provided.
This limitation stems not only from the inclusion of insufficient example quality but also from the absence of explicit guidance, which prevents LLMs from establishing intent-specific mappings between code and comments when relying solely on $<code, comment>$ demonstrations.
In practice, only a subset of code statements is relevant to a given intent, while the remainder is extraneous.
For instance, in Fig. \ref{fig:inference}, only lines 0, 3, and 6 are relevant to the "\textit{What}" intent, whereas the others are irrelevant.
When provided only with \#Code and \#Comment, LLMs struggle to distinguish between relevant and irrelevant information.
Consequently, they may focus on the latter and make incorrect inferences.
To overcome this, we propose enriching demonstration examples with intent-specific important statements that align with their corresponding comments.
These statements serve as a bridge, enabling LLMs to better capture intent-related knowledge and establish precise code–comment correspondences for more accurate comment generation.
Moreover, since different intents emphasize distinct subsets of statements within the same code snippet (e.g., "\textit{What}" and "\textit{Property}" in Fig. \ref{fig:inference}), it is crucial to design a method that can dynamically extract the intent-relevant statements for different intent categories.

\section{Approach}

This section presents a novel approach, \tech{}, for multi-intent code comment generation by optimizing intent-related knowledge for LLMs.
The overall framework of our approach is illustrated in Fig. \ref{fig:overview}, which consists of two main components: 
the \textbf{Example Retriever} and \textbf{Knowledge Augmentation} component.
Given a target code snippet, the \textbf{Example Retriever} first retrieves similar examples (code-comment pairs) from the retrieval corpus.
Then, it assesses the quality of these examples and select the examples containing intent-related knowledge as demonstration examples. 
Subsequently, the \textbf{Knowledge Augmentation} component extracts intent-specific important statements from each demonstration example and integrates them into structured prompts, guiding LLMs to generate accurate intent-specific comments.

\subsection{Example Retriever}

Since LLMs are not specifically trained for multi-intent comment generation, \tech{} adopts the in-context learning strategy.
This subsection describes how demonstration examples are retrieved from the corpus and how their quality is assessed to ensure the inclusion of intent-related knowledge.
To achieve this, we propose a quality assessment mechanism that selects examples with higher semantic consistency between code and comments.
Such high-quality examples not only enhance the reliability of in-context demonstrations but also facilitate the subsequent extraction of intent-specific important statements.

\subsubsection{\textbf{Example Retrieval}} 
In-context learning performs effectively when demonstration examples are carefully selected. 
To leverage the strengths of different retrieval methods, we integrate two common strategies for retrieving similar examples from the corpus whose comments align with the desired intent category. 
These strategies are detailed as follows:

\noindent
\textit{\textbf{Token-based strategy}}: 
This strategy identifies similar code by comparing the overlap of code tokens~\cite{DBLP:conf/wcre/GolubevPPB21,DBLP:conf/icse/GengWD00JML24}. 
It ranks code snippets based on token-level similarity. 
Specifically, it uses Jaccard Coefficient~\cite{niwattanakul2013using} to calculate the token-based similarity $Sim\_Score_{token}$ between a candidate code snippet and the target code. 
The formula is defined as follows:
\vspace{-2pt}
\begin{equation}
\small
	Sim\_Score_{token}=\frac{\mid tokens_{t} \cap tokens_{c} \mid}{\mid tokens_{t} \cup tokens_{c} \mid}
\vspace{-2pt}
\end{equation}
where $tokens_{t}$ and $tokens_{c}$ denote the sub-token lists of the target and candidate code snippets, respectively.

\noindent
\textit{\textbf{Semantic-based strategy}}: 
This strategy focuses on the underlying semantics to retrieve similar code and is widely applied in code clone detection~\cite{DBLP:journals/tosem/ZengYLXWGBDL23,DBLP:journals/corr/abs-2002-08653}. 
In \tech{}, we leverage a pre-trained model~\cite{DBLP:conf/emnlp/ReimersG19} to encode code snippets into embeddings, and then calculate the semantic similarity $Sim\_Score_{semantic}$ using cosine similarity metric:
\begin{equation}
 \resizebox{0.9\columnwidth}{!}{$Sim\_Score_{semantic} = Cosin\_Sim
 (embedding_{t},  embedding_{c})$}
\end{equation}
where $embedding_{t}$ and $embedding_{c}$ denote the embeddings of the target and candidate code snippets.


\subsubsection{\textbf{Examples Quality Assessment}} 
In-context learning requires demonstration examples that not only resemble the target code but also provide intent-related knowledge to guide LLMs, enabling LLMs to extract useful patterns and enhance their inference ability. 
We refer to this property as \textbf{example quality}, which we define as the semantic consistency between a code snippet and its corresponding comment — reflecting the extent to which the example embodies intent-related knowledge.
Following prior work~\cite{DBLP:journals/pacmse/JinL24,DBLP:conf/icse/MastropaoloCPB24}, we quantify this consistency using semantic similarity: higher consistency indicates that the comment is more directly derivable from the code and therefore more beneficial for guiding LLMs. 
Specifically, given a code snippet $X$ and its comment $Y$, we use the encoder $\mathcal{M}_E$ of a pre-trained Encoder-Decoder model~\cite{DBLP:conf/emnlp/WangLGB0H23} to generate  embeddings $\mathcal{M}_E(X)$ and $\mathcal{M}_E(Y)$, respectively.
The example quality score $Quality\_Score$ is then calculated as the cosine similarity between these embeddings:
\vspace{-2pt}
\begin{equation}
\small
\label{eq:example_quality}
 Quality\_Score = Cosin\_Sim(\mathcal{M}_E(X),  \mathcal{M}_E(Y))
\vspace{-2pt}
\end{equation}
A higher $Quality\_Score$ suggests that the example provides stronger guidance, enabling LLMs to identify intent-related knowledge more effectively.
For instance, a code-comment pair with high consistency (Figure \ref{fig:high_code_quality}) allows LLMs to easily establish code–comment correspondences, whereas weakly consistent pairs (Fig.~\ref{fig:low_code_quality}) hinder this process.
In future work, we plan to explore more advanced methods for assessing semantic consistency between code and comments~\cite{DBLP:conf/icse/MastropaoloCPB24}.

In conclusion, during the retrieval phase, we consider both example similarity and example quality.
For a given code snippet, we first use the similarity-based retrieval strategy to retrieve the top-$k$ similar examples $E_1, \dots, E_k$ from the retrieval corpus, each with a similarity score $S_1, \dots, S_k$ and an example quality score $Q_1,\dots, Q_k$, respectively.
To reduce the impact of outliers, we rank these examples on their similarity and example quality scores, denoted as $Sim\_Rank_i$ and $Quality\_Rank_i$, respectively.
We then rank the examples according to the weighted scores (formula (\ref{eq:weigthed_score})).
The top-$f$ examples ($f \leq k$) are then selected as demonstration examples, in line with the in-context learning setting.
\vspace{-2pt}
\begin{equation}
\resizebox{0.9\columnwidth}{!}{$
\label{eq:weigthed_score}
 Example\_Score_i = p * Sim\_Rank_i + (1 - p) * Quality\_Rank_i$}
\vspace{-4pt}
\end{equation}

\subsection{Knowledge Augmentation}

Recent studies~\cite{DBLP:conf/icse/AhmedPDB24v, DBLP:conf/icml/ShrivastavaLT23} have shown that augmenting LLM prompts with information that developers consider during task execution can significantly improve performance on code-related tasks.
When developers tackle a specific-intent code comment generation task, they first identify intent-specific important statements in the code and then formulate the comment based on these statements~\cite{DBLP:conf/icse/MuCSWW23}.
These important statements can also help LLMs construct the intent-specific mapping between the code and comments.
Inspired by this, we enrich demonstration examples with their intent-specific important statements to construct a mapping knowledge chain — linking code to intent-specific statements to comments.
We further leverage CoT reasoning to guide LLMs in (1) extracting intent-specific important statements from the code and (2) generating accurate comments based on these statements, thereby accomplishing multi-intent comment generation tasks.

\subsubsection{\textbf{Statements Extraction}} 
A straightforward solution is to let the LLM perform zero-shot annotation of intent-specific statements for the exemplars and then use them as demonstrations.
However, this creates a circular process in which the LLM first performs statement identification in a zero-shot manner and then learns from its own outputs, effectively reducing the few-shot to zero-shot prompting and amplifying inference errors.
To avoid this circular dependency, we introduce an independent model to identify important statements more reliably.

Building upon prior research~\cite{DBLP:conf/icse/MuCSWW23}, we observed that DOME utilizes an attention mechanism in its generative model to construct a mapping between code and comments based on specific intents. 
This attention mechanism computes an attention matrix that represents the learned mapping.
Leveraging this insight, we propose to \textbf{extract intent-specific important statements using the model's attention matrix.}
However, there are two key limitations in using generative models' attention matrices directly:
(1) \textbf{Data Dependency}: 
Due to the complexity of the tasks, generative models require large amounts of training data to perform effectively, which limits their usefulness due to the scarcity of labelled data.
(2) \textbf{Attention Mistake}: 
Due to the autoregressive mechanism, generative models compute attention based on previously generated tokens~\cite{DBLP:conf/kbse/MuC0WW22}. 
However, if the earlier tokens are predicted incorrectly, subsequent attention computations must be influenced, resulting in an inaccurate attention matrix and hindering the extraction of key statements.
To address these challenges, we propose to leverage a \textbf{code search model} instead. 
Unlike generation tasks, code search is a simpler classification task where the model focuses solely on establishing a mapping between queries and code snippets without the burden of full-generation tasks.
This makes code search models effective even with limited training data. 
Furthermore, since code search models process both code and comments simultaneously, they can compute the correct attention matrix in a single pass.

Therefore, we design a code search task in which, given a query consisting of a comment and its intent, the model identifies the corresponding code from the corpus.
Assume a code snippet $X = [x_1, x_2,\dots,x_N]$ with $N$ tokens and $L$ statements, and its corresponding comment $Y = [y_1, y_2,\dots,y_K]$ with an associated intent $I$.
In this task, following the prior studies~\cite{DBLP:conf/kbse/ZhangP0LG22,DBLP:conf/icse/NiuL0GH022}, we use a cross-encoder~\cite{DBLP:conf/emnlp/FengGTDFGS0LJZ20}, which outperforms dual-encoders\cite{DBLP:conf/iclr/GuoRLFT0ZDSFTDC21} for constructing mappings between queries and code.
Specifically, 
to capture the mapping among comments, intents, and code, 
we concatenate $Y$, $I$ and $X$ into a unified sequence $[Y, I, X] = ( y_1, y_2,\dots,y_K,I,x_1, x_2,\dots,x_L )$ and use a pre-trained model encoder $E(\cdot)$ to map the sequence into an embedding. 
A neural network head $NN(\cdot)$ then maps this embedding to a scalar representing the relevance score between the comment, intent, and code:
\vspace{-2pt}
\begin{equation}
\small
	Relevance\_Score( Y, I, X ) = NN( E( [Y, I, X] ) )
\vspace{-2pt}
\end{equation}
In this manner, the model learns the mapping between code $X$ and comment $Y$ based on the intent $I$.
To obtain this mapping, we draw upon recent research~\cite{DBLP:conf/scam/MohammadkhaniTH23} to select the final attention layer's matrix $A$ from the search model, constructed as follows:
\begin{equation}
  \resizebox{0.9\columnwidth}{!}{$
    A =
    \begin{bmatrix}
      a_{1,1} & \cdots & a_{1,K+1} & a_{1,K+2} & \cdots & a_{1,K+1+N} \\
      \vdots  & \ddots & \vdots    & \vdots    & \ddots & \vdots    \\
      a_{K,1} & \cdots & a_{K,K+1} & a_{K,K+2} & \cdots & a_{K,K+1+N} \\
      a_{K+1,1} & \cdots & a_{K+1,K+1} & a_{K+1,K+2} & \cdots & a_{K+1,K+1+N} \\
      \vdots  & \ddots & \vdots    & \vdots    & \ddots & \vdots    \\
      a_{K+1+N,1} & \cdots & a_{K+1+N,K+1} & a_{K+1+N,K+2} & \cdots & a_{K+1+N,K+1+N}
    \end{bmatrix}
    $
  }
  \label{eq:wideA}
\end{equation}

To obtain the intent-specific important statements from the code, we first need to derive the mapping from each token in the comment to each token in the code, denoted as $A_{Y \mapsto X}$:
\vspace{-2pt}
\begin{equation}
\small
	A_{Y \mapsto X} =
 \begin{bmatrix}
	 a_{1,K+2}   & \cdots & a_{1,K+1+N}   \\
      \vdots      & \ddots & \vdots      \\
      a_{K,K+2}   & \cdots & a_{K,K+1+N}   \\
	 \end{bmatrix}
\vspace{-2pt}
\end{equation}
Then, to obtain the mapping $\widetilde{A}_{Y \mapsto X}$ representing the relationship from the entire comment to each statements in the code, we perform a summation along the $1^{st}$ dimension of $A_{Y \mapsto X}$ and calculate the attention scores from the entire comment to each statement in the code.
\vspace{-2pt}
\begin{equation}
\small
	\widetilde{A}_{Y \mapsto X} =
 \begin{bmatrix}
	 \widetilde{a_1'} & \widetilde{a_2'} & \cdots & \widetilde{a_L'}  
	 \end{bmatrix}
     \vspace{-2pt}
\end{equation}

Finally, we rank the statements according to the scores in $\widetilde{A}_{Y \mapsto X}$ and extract the intent-specific important statements, to augment the demonstration examples.
Building on this, the LLMs can more effectively discern the relationships between code and comments by incorporating these augmented demonstration examples. 
This results in more accurate intent-specific comment generation and superior performance with fewer demonstration examples.

\begin{table}[t]
  \centering
  \small
  \caption{The Example of Prompt of \tech{}.\label{tab:prompt}}
  \label{tab:prompt}%
  \begin{adjustbox}{width=\linewidth,center}
  \begin{threeparttable}
    \begin{tabular}{l|l}
    \toprule
    \textbf{Prompt} & \multicolumn{1}{c}{\textbf{Instantiation}} \\
    \hline
    \makecell[l]{Role \\ Designation} & \makecell[l]{\# You are an expert Java programmer. Give you a code snippet,\\ your task is to \textbf{\textcolor{red}{describe the functionality of}} the code.} \\
    \hline
    \makecell[l]{Chain-of-\\Thought} & \makecell[l]{\# Based on the task itself, some of the statements in the code\\ are more important for you to get the answer, so let's solve the\\ problem step by step:\\Step 1 - extract the important statements from the code, which\\ you should pay more attention to, in order to get the answer.\\Step 2 - \textbf{\textcolor{red}{describe the functionality of}} the code according to\\ the code and the important statements.} \\
    \hline
    \makecell[l]{Example \\ Input \\ \& \\ Output} & \makecell[l]{\# Example Code 1:\\
    \qquad \{Example Code\} \\
    \# Step 1 - Important statements:\\
    \qquad \{Example Important statements\} \\
    \# Step 2 - The comment:\\
    \qquad \{Example Code Comment\}} \\
    \hline
    \makecell[l]{Input} & \makecell[l]{\# For the test code:\\
    \qquad \{Test Code\}} \\
    \hline
    \makecell[l]{Format \\Constraints} & \makecell[l]{\# Please imitate the above example, extract the most important\\ statements of the test code and analyse the code and important\\ statements to use one sentence to \textbf{\textcolor{red}{describe the functionality}}\\ \textbf{\textcolor{red}{of}} the code. Please output the results in the following format:\\\# Step 1 - Important statements:...\\\# Step 2 - The comment:...} \\
    \bottomrule
    \end{tabular}%
    \end{threeparttable}
\end{adjustbox}
\end{table}%

\subsubsection{\textbf{Prompt Construction}}
After obtaining the augmented demonstration examples, we construct the final prompt for LLMs to generate multi-intent comments.
As shown in Table \ref{tab:prompt}, the final prompt consists of five key components: Role Designation, Chain-of-Thought Reasoning, Example Input\&Output, Input, and Format Constraints. 
Each component plays a distinct role, as described below:

\begin{itemize}[leftmargin=*]


\item {\texttt{\textbf{Role Designation}}}: Introduces the multi-intent comment generation task context.

\item {\texttt{\textbf{Chain-of-Thought}}}: Guides LLMs through the multi-intent comment generation process step-by-step.

\item {\texttt{\textbf{Example Input\&Output}}}: Illustrates the task requirements and desired output.

\item {\texttt{\textbf{Input}}}: Represents the specific problem the LLMs need to solve, serving as a practical evaluation scenario.

\item {\texttt{\textbf{Format Constraints}}}: Defines the expected output format for LLMs.
\end{itemize}
\vspace{-2pt}

These components are concatenated to construct the final prompt, where the desired comment intent (highlighted in red bold) follows definitions from prior work~\cite{DBLP:conf/icse/GengWD00JML24}.
Specifically, the instructions are: “describe the functionality of” (\textit{What}), “explain the reason why the method is provided or the design rationale of” (\textit{Why}), “describe the usage or the expected set-up of using” (\textit{How-to-use}), “describe the implementation details of” (\textit{How-it-is-done}), and “describe the asserted properties of the code, including pre-conditions or post-conditions of” (\textit{Property}).
The LLMs then generate intent-specific comments under the "The comment:" field.

\section{Experiment Design}
\label{sec:exmperiment}


In the following experimental evaluation, we aim to answer the following research questions:
\begin{itemize}[leftmargin=*]
\item \textbf{RQ1}: How does \tech{} perform compared to the state-of-the-art baselines?
\item \textbf{RQ2}: How does each component in \tech{} contribute to the overall performance?
\item \textbf{RQ3}: How well is our code search task in \tech{}?
\item \textbf{RQ4}: How does \tech{} perform in human evaluation?
\end{itemize}





\begin{table}[t]
\small
  \centering

  \caption{Statistic of Funcom and TLC datasets}
  \begin{adjustbox}{width=0.48\textwidth,center}
 \begin{threeparttable}
    \begin{tabular}{c|r|r|r|r|r|r}
    \toprule
    \multirow{2}{*}{Category} & \multicolumn{3}{c|}{Funcom} & \multicolumn{3}{c}{TLC} \\
\cmidrule{2-7}          & \multicolumn{1}{c|}{Train} & \multicolumn{1}{c|}{Test} & \multicolumn{1}{c|}{Validation}  & \multicolumn{1}{c|}{Train} & \multicolumn{1}{c|}{Test} & \multicolumn{1}{c}{Validation} \\
    \midrule
    What  & 684121 & 37399 & 36525 & 29654 & 2773 & 4177 \\
    Why   & 152320 & 7403 & 8317  & 6219  & 600 & 889 \\
    How-to-use & 24862 & 1085 & 1403  & 886   & 72 & 127 \\
    How-it-is-done & 150432 & 7281 & 7585  & 11698 & 1089 & 1605 \\
    Property & 167188 & 8070 & 8553  & 5071  & 451 & 757 \\
    \midrule
    Total & 1178923 & 61238 & 62383 & 53528 & 4985 & 7555 \\
    \bottomrule
    \end{tabular}%
    \end{threeparttable}
\end{adjustbox}
  \label{tab:dataset}%
\end{table}%

\begin{table*}[t]
\centering
\tiny
\belowrulesep=0pt
\aboverulesep=0pt
\caption{Performances of \tech{} and baselines on average\label{tab:rq1_average}}
 \begin{adjustbox}{width=1\textwidth,center}
 \begin{threeparttable}
    \begin{tabular}{c|l|l|l|rrrr|rrrr}
    \toprule
    \multirow{2}{*}{LLM} & \multicolumn{3}{c|}{\multirow{2}{*}{Method}} & \multicolumn{4}{c|}{TLC}      & \multicolumn{4}{c}{Funcom} \\
\cmidrule{5-12}          & \multicolumn{3}{c|}{} & BLEU  & METEOR & ROUGE-L & SBERT & BLEU  & METEOR & ROUGE-L & SBERT \\
    \midrule
    Transformer & \multicolumn{3}{l|}{Dome} & 22.12  & 19.24  & 37.08  & 51.52  & 23.10  & 17.93  & 30.55  & 49.39  \\
    \midrule
    \multirow{10}{*}{CodeLlama} & \multicolumn{1}{l|}{\multirow{5}{*}{FSMIC}} & \multicolumn{2}{l|}{0-shot} & 12.25  & 17.28  & 29.62  & 56.16  & 9.40  & 2.97  & 8.92  & 6.70  \\
\cmidrule{3-12}          &       & \multicolumn{1}{l|}{\multirow{2}{*}{3-shot}} & semantic & 22.63  & 25.55  & 41.75  & 62.33  & 17.67  & 19.33  & 30.92  & 55.38  \\
\cmidrule{4-4}          &       &       & token & 24.20  & 25.86  & 43.03  & 62.15  & 16.34  & 19.52  & 30.39  & 55.41  \\
\cmidrule{3-12}          &       & \multicolumn{1}{l|}{\multirow{2}{*}{5-shot}} & semantic & 23.38  & 26.01  & 42.41  & 62.63  & 18.79  & 20.47  & 32.81  & 56.53  \\
\cmidrule{4-4}          &       &       & token & 23.07  & 24.60  & 41.27  & 59.94  & 17.60  & 20.00  & 32.07  & 56.49  \\

\cmidrule{2-12}          & \multicolumn{1}{l|}{\multirow{5}{*}{\tech{}}} & \multicolumn{2}{l|}{0-shot} & 19.77  & 18.93  & 36.82  & 59.64  & 21.25  & 16.52  & 29.64  & 54.93  \\
\cmidrule{3-12}          &       & \multicolumn{1}{l|}{\multirow{2}{*}{3-shot}} & semantic & {29.13} & {28.52} & {47.54} & {66.82} & {25.20} & {22.22} & {37.47} & {59.37} \\
\cmidrule{4-4}          &       &       & token & {29.78} & {29.47} & {48.64} & {67.54} & {25.11} & {21.98} & {37.25} & {59.33} \\
\cmidrule{3-12}          &       & \multicolumn{1}{l|}{\multirow{2}{*}{5-shot}} & semantic & 29.51  & 29.09  & 47.99  & 66.87  & 25.93  & 22.86  & 38.63  & 59.91  \\
\cmidrule{4-4}          &       &       & token & 30.13  & 29.47  & 48.96  & 67.41  & 25.92  & 22.80  & 38.52  & 59.99  \\
    \midrule
    \multirow{10}{*}{Llama3} & \multicolumn{1}{l|}{\multirow{5}{*}{FSMIC}} & \multicolumn{2}{l|}{0-shot} & 12.65  & 18.90  & 31.36  & 55.67  & 9.47  & 4.79  & 11.56  & 6.86  \\
\cmidrule{3-12}          &       & \multicolumn{1}{l|}{\multirow{2}{*}{3-shot}} & semantic & 16.09  & 18.54  & 32.07  & 51.26  & 14.26  & 14.87  & 23.45  & 42.09  \\
\cmidrule{4-4}          &       &       & token & 16.58  & 17.52  & 31.50  & 46.06  & 14.29  & 13.67  & 21.89  & 37.60  \\
\cmidrule{3-12}          &       & \multicolumn{1}{l|}{\multirow{2}{*}{5-shot}} & semantic & 14.67  & 17.02  & 29.61  & 48.66  & 13.68  & 14.19  & 22.08  & 40.94  \\
\cmidrule{4-4}          &       &       & token & 16.61  & 16.74  & 30.85  & 45.60  & 14.35  & 12.45  & 20.25  & 35.65  \\

\cmidrule{2-12}          & \multicolumn{1}{l|}{\multirow{5}{*}{\tech{}}} & \multicolumn{2}{l|}{0-shot} & 19.02  & 20.84  & 38.95  & 60.11  & 19.91  & 19.71  & 33.84  & 55.83  \\
\cmidrule{3-12}          &       & \multicolumn{1}{l|}{\multirow{2}{*}{3-shot}} & semantic & {29.39} & {29.00} & {49.40} & {66.72} & {24.70} & {23.26} & {39.56} & {59.38} \\
\cmidrule{4-4}          &       &       & token & {29.69} & {29.40} & {49.49} & {66.99} & {24.69} & {23.55} & {39.69} & {59.47} \\
\cmidrule{3-12}          &       & \multicolumn{1}{l|}{\multirow{2}{*}{5-shot}} & semantic & 29.59  & 29.25  & 49.66  & 66.96  & 25.23  & 23.77  & 40.41  & 59.74  \\
\cmidrule{4-4}          &       &       & token & 30.03  & 29.69  & 49.77  & 67.12  & 25.50  & 24.05  & 40.54  & 59.98  \\

    \midrule
    \multirow{10}{*}{Qwen2.5-Coder} & \multicolumn{1}{l|}{\multirow{5}{*}{FSMIC}} & \multicolumn{2}{l|}{0-shot} & 6.57 & 3.28 & 7.88 & 8.21 & 6.72 & 3.43 & 7.98 & 8.22  \\
\cmidrule{3-12}          &       & \multicolumn{1}{l|}{\multirow{2}{*}{3-shot}} & semantic & 18.46 & 22.23 & 34.1 & 60.42 & 14.38 & 19.02 & 29.03 & 54.40  \\
\cmidrule{4-4}          &       &       & token & 19.84 & 22.65 & 35.41 & 61.39 & 15.09 & 19.42 & 29.85 & 54.98  \\
\cmidrule{3-12}          &       & \multicolumn{1}{l|}{\multirow{2}{*}{5-shot}} & semantic & 18.93 & 22.52 & 34.61 & 60.87 & 14.97 & 19.44 & 29.80 & 54.93  \\
\cmidrule{4-4}          &       &       & token & 20.35 & 22.99 & 36.07 & 61.78 & 15.75 & 19.82 & 30.64 & 55.50  \\

\cmidrule{2-12}          & \multicolumn{1}{l|}{\multirow{5}{*}{\tech{}}} & \multicolumn{2}{l|}{0-shot} & 13.51 & 17.65 & 26.97 & 55.13 & 13.84 & 17.02 & 26.05 & 52.16  \\
\cmidrule{3-12}          &       & \multicolumn{1}{l|}{\multirow{2}{*}{3-shot}} & semantic & {24.63} & {24.73} & {39.96} & {63.72} & {20.90} & {21.63} & {35.12} & {57.79} \\
\cmidrule{4-4}          &       &       & token & {25.70} & {25.28} & {41.19} & {64.69} & {21.22} & {21.93} & {35.49} & {58.17} \\
\cmidrule{3-12}          &       & \multicolumn{1}{l|}{\multirow{2}{*}{5-shot}} & semantic & 25.29 & 25.10 & 40.77 & 64.33 & 21.66 & 22.11 & 35.96 & 58.32  \\
\cmidrule{4-4}          &       &       & token & 26.26 & 25.66 & 41.98 & 65.10 & 22.05 & 22.47 & 36.43 & 58.73  \\

    \bottomrule
\end{tabular}

\end{threeparttable}
\end{adjustbox}
\end{table*}

\subsection{Dataset}

In this work, we use the multi-intent comment generation datasets from previous studies~\cite{DBLP:conf/icse/MuCSWW23, DBLP:conf/icse/GengWD00JML24} for evaluation. 
In concrete, we employ two widely used Java datasets: Funcom~\cite{DBLP:conf/icse/LeClairJM19} and TLC~\cite{DBLP:conf/ijcai/HuLXLLJ18}.
Funcom contains 2.1M code-comment pairs from 29K Java projects, originally collected by Lopes et al.~\cite{funcom/url} and further cleaned by LeClair et al.~\cite{DBLP:conf/icse/LeClairJM19}. 
TLC comprises 87,136 code-comment pairs collected from over 9K Java projects with at least 20 stars. 
The intent categories of comments in these datasets were labeled by DOME~\cite{DBLP:conf/icse/MuCSWW23}. 
They recruited five domain experts to label the intents of 8K code snippets manually and then fine-tuned a model as a classifier on this labeled data. 
DOME then applied this fine-tuned model to predict the intent category for each comment, using these predictions as ground-truth labels.
Following the previous work~\cite{DBLP:conf/icse/GengWD00JML24}, we serve the training sets as the retrieval corpus. 
To alleviate data leakage between the training sets and test sets (e.g., LLMs directly "copying"  comments from the same source), we identify and remove duplicate comments between the training and test sets. 
The dataset statistics are shown in Table \ref{tab:dataset}. 
This dataset is also used for our code search task.

\subsection{Evaluation Metrics}

We evaluate the performance of \tech{} and the baselines using widely adopted metrics, including BLEU~\cite{DBLP:conf/acl/PapineniRWZ02}, ROUGE-L~\cite{Lin_2004}, METEOR~\cite{DBLP:conf/acl/BanerjeeL05}, and SBERT~\cite{DBLP:conf/iwpc/HaqueEBM22}. 
BLEU (BiLingual Evaluation Understudy) is a precision-based metric computing the weighted geometric mean of modified 1- to 4-gram precisions, with length penalty. 
In this study, we use BLEU-4, which is the standard choice in code comment generation.
METEOR improves upon BLEU by combining n-gram precision and recall through their harmonic mean, providing a more balanced similarity measure.
ROUGE-L measures the longest common subsequence (LCS) between the generated and reference comments, focusing on recall and content coverage.
Unlike these three metrics, which focus on textual similarity, SBERT ~\cite{DBLP:conf/emnlp/ReimersG19} assesses semantic similarity by computing cosine similarity between embeddings of generated and reference comments.
Higher values across these metrics indicate closer alignment with ground-truth comments, reflecting stronger performance in code summarization.

\subsection{Experimental Settings}

\noindent
\textbf{\textit{Base LLMs.}}
We conduct experiments on three popular autoregressive LLMs: CodeLlama~\cite{DBLP:journals/corr/abs-2307-09288}, Llama3~\cite{llama3/url} and Qwen2.5-Coder~\cite{qwen2025qwen25technicalreport}.
CodeLlama, developed by Meta and based on Llama 2, supports infilling, large input contexts, and zero-shot instruction-following programming tasks. 
Llama-3, the latest iteration in the Llama series by Meta, provides enhanced processing power and versatility. 
Qwen2.5-Coder, a code-specialized model from Alibaba’s Qwen 2.5 series, achieves state-of-the-art performance in programming tasks among open models. 
In our experiment, we use CodeLlama-34B, Llama-3-70B, and Qwen2.5-Coder-32B, setting the temperature to 0.5 to produce well-defined outputs, following prior work~\cite{DBLP:conf/icse/GengWD00JML24}. 

\smallskip
\noindent
\textbf{\textit{Baselines.}}
The baselines for comparison include DOME~\cite{DBLP:conf/icse/MuCSWW23}, the first approach designed for multi-intent comment generation, and FSMIC, proposed by Geng et al.~\cite{DBLP:conf/icse/GengWD00JML24}, which leverages LLMs with few-shot learning to generate intent-specific comments.
We exclude the work of Sun et al.\cite{DBLP:conf/icse/SunMLZFLDLC25}, as their method is essentially identical to FSMIC.
To ensure a fair comparison and isolate methodological differences, both \tech{} and FSMIC used the same LLM in their implementations.

\smallskip
\noindent
\textbf{\textit{Implementation.}}
All experiments were conducted on a server with four NVIDIA A800 GPUs and 512 GB of memory, running Ubuntu 20.04.2.
Following prior work~\cite{DBLP:conf/icse/GengWD00JML24}, we evaluate \tech{} and FSMIC under zero-shot and few-shot settings, using 3 and 5 demonstrations.
For the code search task, we adopt CodeBERT~\cite{DBLP:conf/emnlp/FengGTDFGS0LJZ20}, trained for 10 epochs with a learning rate of 2e-5 (linear decay, no weight decay), consistent with the CodeBERT series.
To ensure scalability, we follow prior studies~\cite{DBLP:conf/issta/GuoLWL0HZ023,DBLP:journals/corr/abs-1708-05031,DBLP:conf/emnlp/FengGTDFGS0LJZ20,DBLP:conf/emnlp/ZhouGLZSDLMWD22,DBLP:conf/sigir/FormalLPC22} in pairing fewer negative samples with each positive instance as the dataset size increases.
Specifically, consistent with prior work~\cite{DBLP:journals/corr/abs-1708-05031,DBLP:conf/emnlp/ZhouGLZSDLMWD22}, we used 5 negatives per positive for Funcom and 20 for TLC. 
For each comment in a positive instance, where negatives were constructed by randomly sampling intent–code pairs from the training set.
In Example Quality Assessment, $k$ and $p$ were set to 10 and 0.8, respectively, ensuring that similarity dominates the $Example\_Score$.
To mitigate randomness in the LLM inference, each experiment was repeated five times, and the results were averaged.



\section{Result}

\subsection{RQ1: Comparison with Baselines}
\label{sec:rq1}

\smallskip 
\noindent 
\textbf{Setups.}
As detailed in Section \ref{sec:exmperiment}, we evaluate \tech{} and two state-of-the-art baselines, DOME and FSMIC, on the multi-intent comment generation task using four widely adopted metrics: BLEU, ROUGE-L, METEOR, and SBERT.
We conducted experiments with both semantic-based and token-based demonstration selection strategies, calculating the average performance for each method. 
Furthermore, to evaluate comment generation effectiveness across different intents, we split the test dataset into five groups based on intent categories and measured the average performance for each intent.

\smallskip
\noindent
\textbf{Results.}
Table \ref{tab:rq1_average} compares the average performance of \tech{} against the baselines.  
As shown in Table \ref{tab:rq1_average}, \tech{} consistently outperforms across nearly all metrics, regardless of whether token-based or semantic-based selection is used.
As anticipated, LLM performance improves with more demonstrations in few-shot learning. 
Interestingly, FSMIC with 3-shot surpasses its 5-shot version, likely due to potential contamination from intent-agnostic examples, whereas \tech{} with 3-shot slightly underperforms its 5-shot variant but greatly reduces computational cost.
To emphasize the efficiency of our method, we adopt 3-shot for \tech{} and 5-shot for FSMIC in subsequent experiments, where \tech{} continues to outperform FSMIC despite using fewer demonstrations.
\uline{Compared to the best baselines, \tech{} improves the average performance of BLEU, METEOR, ROUGE-L, and SBERT by 14.49\%, 22.41\%, 20.72\%, and 12.94\% across both datasets.}
The Mann-Whitney U test~\cite{DBLP:conf/uss/0008ZL0C23}  confirms the statistical significance of these improvements, with all p-values less than 9.51e-4.
For instance, on CodeLlama, \tech{}-3-shot with semantic-based selection improves BLEU, METEOR, ROUGE-L, and SBERT by 24.59\%, 9.65\%, 12.10\%, and 6.69\% across both datasets, respectively, compared to the best baseline. 
For the non-code model Llama3, although its overall few-shot performance is lower than the other two code-specific models, \tech{} achieves comparable or even higher performance, demonstrating its ability to enhance the model’s understanding of code comment generation.
However, when applied to Qwen2.5-Coder, the performance gains are less pronounced, and on the Funcom dataset, the BLEU score is slightly lower than that of the baseline.
This can be attributed to differences in tokenization and lexical usage between the Qwen series and the Llama family.
Specifically, Qwen2.5-Coder often generates semantically equivalent yet lexically divergent tokens, which negatively affect token-level metrics such as BLEU, while metrics emphasizing semantic similarity still capture the improvements.

We further analyzed the average performance across the five intent categories, as presented in Table \ref{tab:rq1_intents}.
Due to space constraints, we only report the results for DOME, FSMIC-5-shot, and \tech{}-3-shot on CodeLlama; full results are available in our repository.
As shown in Table \ref{tab:rq1_intents}, \uline{\tech{} consistently outperforms all baselines across each intent category.}
In particular, \tech{} achieves substantial improvements on the “\textit{What}” and “\textit{Why}” intents under the METEOR metric, while gains are more moderate for “\textit{How-to-use}”.
This is attributed to \tech{}'s enhanced capability in mapping complex code-comment patterns, whereas simpler intents already benefit adequately from standard few-shot examples.
We also observed distinct advantages across different retrieval strategies.
For instance, token-based selection performs better on TLC, where shorter code snippets favor lexical matching,
while semantics-based selection excels on Funcom, as longer code requires deeper semantic analysis.
These findings highlight the importance of matching the retrieval strategy to the corpus characteristics.

\begin{table}[t]
\small
\centering
\belowrulesep=0pt
\aboverulesep=0pt
\caption{Performances of \tech{} and baselines on each intent category on CodeLlama\label{tab:rq1_intents}}

 \begin{adjustbox}{width=0.49\textwidth,center}
 \begin{threeparttable}
    \begin{tabular}{c|l|l|cccc}
    \toprule
    Intent & \multicolumn{2}{c|}{Method} & BLEU  & METEOR & ROUGE-L & SBERT  \\
    \midrule
    \multirow{5}{*}{What} & \multicolumn{2}{l|}{DOME} & 23.85  & 20.10  & 38.29  & 52.54    \\
\cmidrule{2-7}          & \multicolumn{1}{l|}{\multirow{2}{*}{FSMIC}} & semantic & 23.21  & 26.68  & 42.57  & 61.75  \\
\cmidrule{3-3}          &       & token & 22.41  & 24.63  & 40.67  & 57.94    \\
\cmidrule{2-7}          & \multicolumn{1}{l|}{\multirow{2}{*}{\tech{}}} & semantic & 31.23  & 30.18  & 49.34  & 67.25   \\
\cmidrule{3-3}          &       & token & \textbf{32.08} & \textbf{30.94} & \textbf{50.60} & \textbf{68.26}  \\
    \midrule
    \multirow{5}{*}{Why} & \multicolumn{2}{l|}{DOME} & 20.51  & 16.92  & 35.13  & 46.90    \\
\cmidrule{2-7}          & \multicolumn{1}{l|}{\multirow{2}{*}{FSMIC}} & semantic & 22.89  & 22.64  & 38.78  & 58.78  \\
\cmidrule{3-3}          &       & token & 22.40  & 22.49  & 39.31  & 59.48    \\
\cmidrule{2-7}          & \multicolumn{1}{l|}{\multirow{2}{*}{\tech{}}} & semantic & 26.73  & 25.49  & 43.95  & 63.67    \\
\cmidrule{3-3}          &       & token & \textbf{27.06} & \textbf{27.01} & \textbf{44.21} & \textbf{64.15}  \\
    \midrule
    \multirow{5}{*}{How-to-use} & \multicolumn{2}{l|}{DOME} & 14.66  & 12.68  & 24.78  & 38.51    \\
\cmidrule{2-7}          & \multicolumn{1}{l|}{\multirow{2}{*}{FSMIC}} & semantic & 14.40  & 14.09  & 25.10  & 47.85 \\
\cmidrule{3-3}          &       & token & 14.67  & 14.32  & 25.86  & 46.88    \\
\cmidrule{2-7}          & \multicolumn{1}{l|}{\multirow{2}{*}{\tech{}}} & semantic & 17.00  & 15.17  & 26.67  & 50.39    \\
\cmidrule{3-3}          &       & token & \textbf{17.40} & \textbf{16.61} & \textbf{29.73} & \textbf{52.37} \\
    \midrule
    \multirow{5}{*}{How-it-is-done} & \multicolumn{2}{l|}{DOME} & 19.05  & 17.97  & 34.94  & 50.38  \\
\cmidrule{2-7}          & \multicolumn{1}{l|}{\multirow{2}{*}{FSMIC}} & semantic & 22.10  & 25.38  & 42.14  & 65.30  \\
\cmidrule{3-3}          &       & token & 22.18  & 23.87  & 40.74  & 61.78    \\
\cmidrule{2-7}          & \multicolumn{1}{l|}{\multirow{2}{*}{\tech{}}} & semantic & 24.68  & 26.30  & 44.60  & 65.91    \\
\cmidrule{3-3}          &       & token & \textbf{25.54} & \textbf{26.87} & \textbf{45.05} & \textbf{67.24}  \\
    \midrule
    \multirow{5}{*}{Property} & \multicolumn{2}{l|}{DOME} & 22.26  & 21.09  & 38.77  & 56.21   \\
\cmidrule{2-7}          & \multicolumn{1}{l|}{\multirow{2}{*}{FSMIC}} & semantic & 28.27  & 30.91  & 49.54  & 70.37    \\
\cmidrule{3-3}          &       & token & 30.95  & 31.98  & 51.98  & 70.36    \\
\cmidrule{2-7}          & \multicolumn{1}{l|}{\multirow{2}{*}{\tech{}}} & semantic & 31.98  & 30.99  & 52.33  & 71.24   \\
\cmidrule{3-3}          &       & token & \textbf{32.97} & \textbf{32.96} & \textbf{53.61} & \textbf{71.98}  \\
    \bottomrule
\end{tabular}

\end{threeparttable}
\end{adjustbox}
\end{table}

To further evaluate the performance of \tech{} on more advanced LLMs, we conducted additional experiments on the proprietary GPT-3.5-turbo-instruct, which offers a cost advantage of approximately 30$\times$ compared to GPT-4.
Specifically, we randomly sampled 20 test cases from each intent category across both datasets (200 in total) and compared \tech{}-3-shot with FSMIC-5-shot under semantic-based selection.
As shown in Table \ref{tab:rq3}, \uline{\tech{} consistently outperforms FSMIC across both datasets, indicating that the knowledge-augmentation remains effective even with a high-performing proprietary model.}
This suggests that knowledge augmentation enhances LLM reasoning by providing clearer inference guidance, independent of model capacity.
Moreover, we observed that GPT-3.5-turbo-instruct often failed to answer the questions when used with few-shot learning alone.
These findings indicate that beyond controlling the inference process of LLMs, \tech{} also provides clearer formulations of tasks, improving comprehension and enabling the generation of more accurate and relevant outputs.


\subsection{RQ2: Ablation Experiment}

\smallskip 
\noindent 
\textbf{Setups.}
We conducted an ablation study to evaluate the contribution of each core component in \tech{}, comparing it with five variants:
(1) \textit{w/o} CoT: Removes CoT while retaining Knowledge Augmentation, to evaluate CoT's role in enhancing the effectiveness of Knowledge Augmentation.
(2) \textit{w/o} KA: Omits Knowledge Augmentation (KA) while retaining CoT during prompt construction, to assess the importance of augmenting prompts with additional context.
(3) \textit{w/o} EQ: Excludes Example Quality Assessment (EQ) during the example retrieval phase, to assess the significance of selecting high-quality demonstration examples.
(4) \textit{w/o} KA\&EQ: Removes both KA and EQ, retaining only CoT, to isolate CoT’s impact.
(5) \textit{w/o} CS: Uses the model from DOME instead of our code search model for extracting intent-specific important statements during Knowledge Augmentation, to assess the impact of model difference.
We reuse the evaluation metrics from RQ1 for comparison.

\smallskip 
\noindent 
\textbf{Results.}
Table \ref{tab:rq2} presents the ablation results for CodeLlama (additional results are in our repository).
As shown in Table \ref{tab:rq2}, \uline{each component contributes positively to \tech{}’s overall performance of \tech{}.}
Specifically, the performance drop in \tech{} \textit{w/o} CoT indicates that augmenting the knowledge in demonstration examples is crucial for guiding LLMs; without it, LLMs cannot effectively leverage the available information. 
Similarly, the lower performance of \tech{} \textit{w/o} KA compared to \tech{} suggests that only guiding LLMs to identify the intent-related knowledge, without augmenting the demonstration examples, may lead to the extraction of incorrect or noisy information, thereby reducing overall performance.
The gap between \tech{} \textit{w/o} EQ and \tech{} underscores the importance of selecting semantically consistent examples to ensure intent-related knowledge.
When both KA and EQ are removed, performance drops sharply, approaching the level of FSMIC in Table \ref{tab:rq1_average}, showing that CoT alone provides limited benefit compared to KA and EQ.
Finally, the comparison between \tech{} \textit{w/o} CS and \tech{} demonstrates that the statements extracted by our code search model are more effective in helping LLMs construct accurate mappings among intents, code, and comments.

\begin{table}[t]

\centering
\belowrulesep=0pt
\aboverulesep=0pt
\caption{Performances of \tech{} on GPT-3.5-turbo-instruct\label{tab:rq3}}
 \begin{adjustbox}{width=0.48\textwidth,center}
 \begin{threeparttable}

\begin{tabular}{l|c|cccc}
    \toprule
    \multicolumn{1}{c|}{Dataset} & \multicolumn{1}{c|}{Method} & \multicolumn{1}{c}{BLEU} & \multicolumn{1}{c}{METEOR} & \multicolumn{1}{c}{ROUGE-L} & \multicolumn{1}{c}{SBERT} \\
    \midrule
    \multirow{2}{*}{TLC} & FSMIC & 24.38 & 29.65 & 37.06 & 59.58 \\
          & IGMIC & \textbf{31.50} & \textbf{34.29} & \textbf{46.20} & \textbf{72.37} \\
    \midrule
    \multirow{2}{*}{Funcom} & FSMIC & 31.32 & 35.14 & 47.96 & 68.96 \\
          & IGMIC & \textbf{33.00} & \textbf{35.30} & \textbf{50.73} &\textbf{69.77} \\
    \bottomrule
    \end{tabular}%

\end{threeparttable}
\end{adjustbox}
\end{table}

\begin{table*}[t!]
\centering
\small
\belowrulesep=0pt
\aboverulesep=0pt
\caption{Ablation Experiment Results on CodeLlama\label{tab:rq2}}
 \begin{adjustbox}{width=0.8\textwidth,center}
 \begin{threeparttable}

    \begin{tabular}{ll|cccc|cccc}
    \toprule
    \multicolumn{2}{c|}{\multirow{2}{*}{Method}} & \multicolumn{4}{c|}{TLC}      & \multicolumn{4}{c}{Funcom} \\
\cmidrule{3-10} &   & BLEU  & METEOR & ROUGE-L & SBERT & BLEU  & METEOR & ROUGE-L & \multicolumn{1}{c}{SBERT} \\
    \midrule
    \multicolumn{1}{l|}{\multirow{2}{*}{\tech{}}} & semantic & \textbf{29.13} & \textbf{28.52} & \textbf{47.54} & \textbf{66.82} & \textbf{25.20} & \textbf{22.22} & \textbf{37.47} & \textbf{59.37} \\
\cmidrule{2-2}    \multicolumn{1}{l|}{} & token & \textbf{29.78} & \textbf{29.47} & \textbf{48.64} & \textbf{67.54} & \textbf{25.11} & \textbf{21.98} & \textbf{37.25} & \textbf{59.33} \\
    \midrule
    \multicolumn{1}{l|}{\multirow{2}{*}{\tech{} \textit{w/o} CoT}} & semantic & 16.05  & 21.01  & 35.19  & 57.02  & 14.54  & 17.44  & 27.29  & 48.37 \\
\cmidrule{2-2}    \multicolumn{1}{l|}{} & token & 17.44  & 21.20  & 36.04  & 55.54  & 14.32  & 16.98 & 26.72  & 46.54  \\

\midrule
    \multicolumn{1}{l|}{\multirow{2}{*}{\tech{} w/o KA}} & semantic & 24.73  & 25.43  & 43.97  & 63.76  & 22.15  & 19.95  & 33.77  & 55.99  \\
\cmidrule{2-2}    \multicolumn{1}{l|}{} & token & 25.90  & 26.44  & 45.19  & 65.02  & 22.33  & 19.89  & 33.78  & 55.51  \\
\midrule
    \multicolumn{1}{l|}{\multirow{2}{*}{\tech{} \textit{w/o} EQ}} & semantic & 26.92  & 27.11  & 45.39  & 65.66  & 23.32  & 20.91  & 35.33  & 58.47 \\
\cmidrule{2-2}    \multicolumn{1}{l|}{} & token & 28.24  & 28.29  & 46.58  & 66.71  & 23.79  & 21.18  & 35.78  & 58.80  \\

\midrule
\multicolumn{1}{l|}{\multirow{2}{*}{\tech{} \textit{w/o} KA\&EQ}} & semantic & 24.14 & 25.37 & 38.58 & 61.84  & 22.17 & 19.94 & 33.78 & 55.96 \\
\cmidrule{2-2}    \multicolumn{1}{l|}{} & token & 25.29 & 24.24 & 40.20 & 63.36  & 22.60 & 20.00 & 34.02 & 55.71  \\

\midrule
\multicolumn{1}{l|}{\multirow{2}{*}{\tech{} \textit{w/o} CS}} & semantic & 28.02  & 25.30  & 42.76  & 65.00  & 23.25  & 20.94  & 34.94  & 58.07  \\
\cmidrule{2-2}    \multicolumn{1}{l|}{} & token & 28.67  & 25.84  & 43.58  & 65.73  & 23.23  & 21.01  & 34.98  & 58.27  \\
    \bottomrule
    \end{tabular}%

\end{threeparttable}
\end{adjustbox}
\end{table*}

\subsection{RQ3: Advantage of Search Task}

\smallskip 
\noindent 
\textbf{Setups.}
The ablation experiment has demonstrated that the important statements extracted by our code search model outperform those generated by the generative model. 
In the above experiments, we trained all models on the entire TLC and Funcom datasets to ensure robust learning of intent-specific mappings between code and comments.
To further highlight the lightweight nature of our search model, we evaluated its performance against the generative model under limited training data conditions. 
Specifically, we trained the CodeLlama on different proportions of the TLC and Funcom datasets, resulting in six variants for comparison.

\smallskip 
\noindent 
\textbf{Results.}
Due to space constraints, we only report the results for TLC in Table \ref{tab:discuss-data}.
As the amount of training data decreases, the performance of \tech{} \textit{w/o} CS drops substantially, while \tech{} remains stable. 
Notably, with only 25\% of the training data, \tech{} outperforms \tech{} \textit{w/o} CS trained with 75\% of the data.
This indicates that generative models, which handle complex tasks, are highly data-dependent. 
In contrast, the code search model benefits from its dataset construction mechanism, which produces abundant high-quality training data even under limited resources.
Thus, \uline{\tech{} with the lightweight search model maintains robust performance under data scarcity}, effectively capturing intent-specific mappings.

\begin{table}[t]
\belowrulesep=0pt
\aboverulesep=0pt
\centering
\caption{Performances of \tech{} and \tech{} \textit{\textit{w/o}} CS on different scale of training data of TLC\label{tab:discuss-data}}
 \begin{adjustbox}{width=1\linewidth,center}
 \begin{threeparttable}

\begin{tabular}{l|cccc}
    \toprule
    \multicolumn{1}{c|}{\multirow{1}{*}{Method}}       & BLEU  & METEOR & ROUGE-L & SBERT  \\
    \midrule
    \makecell[l]{\ \ 75\% data - \tech{}} &  27.74  & 25.46  & 42.34  & 65.00   \\
    \makecell[l]{\ \ 75\% data - \tech{} \textit{w/o} CS} &  25.55  & 24.06  & 40.14  & 63.95   \\
    \makecell[l]{\ \ 50\% data - \tech{}}   & 27.85  & 25.49  & 42.52 & 64.98    \\
    \makecell[l]{\ \ 50\% data - \tech{} \textit{w/o} CS}   & 22.81  & 22.62  & 37.08 & 62.74    \\
    \makecell[l]{\ \ 25\% data - \tech{}}   & 27.81  & 25.35  & 42.39 & 64.82    \\
    \makecell[l]{\ \ 25\% data - \tech{} \textit{w/o} CS}   & 20.16  & 20.92  & 34.12 & 61.33   \\
    \bottomrule
    \end{tabular}%

\begin{tablenotes}
  \small
\item[*]
All variants use 3-shot learning with semantic-based selection, and the training data is randomly sampled from the training set.

\end{tablenotes}

\end{threeparttable}
\end{adjustbox}
\end{table}

\subsection{RQ4: Human Evaluation}

\smallskip 
\noindent 
\textbf{Setups.}
Although current four evaluation metrics can measure the lexical gaps between generated comments and ground truth, they often fail to capture semantic differences. 
To address this, we conducted a human evaluation.
Following prior works~\cite{DBLP:conf/icse/MuCSWW23, DBLP:conf/icse/GengWD00JML24}, we randomly selected 100 code snippets from both datasets (20 per intent category, averaging 12 lines per snippet and 12 words per comment).
For each snippet, participants evaluated comments generated by DOME, FSMIC, and \tech{}, producing 300 generated comments in total.
Then, we recruited six participants with at least six years of Java development experience, including one Ph.D. student and five Master's students who are not co-authors of this paper. 
To ensure fairness, the participants were not informed of the source of each comment, and given five days to complete this task.
Each participant was asked to rate all 300 comments on four aspects:
(1) \textbf{Accuracy}, which reflects the correctness of the generated comment in terms of its consistency with the code;
(2) \textbf{Adequacy}, which reflects the richness of information provided in the generated comments;
(3) \textbf{Intention,} which reflects whether the comments clearly express the target intent;
(4) \textbf{Naturalness}, which reflects the fluency of generated comments from the perspective of grammar.
Participants used a 5-point Likert scale to rate each aspect (1 for poor, 2 for marginal, 3 for acceptable, 4 for good, and 5 for excellent).
This experimental setup is consistent with existing studies~\cite{DBLP:conf/icse/GengWD00JML24,DBLP:conf/icse/MuCSWW23,DBLP:conf/sigsoft/RoyFA21,DBLP:conf/kbse/MuC0WW22}.

\smallskip 
\noindent 
\textbf{Results.}
Table~\ref{tab:rq4} presents the human evaluation results.
The Fleiss’s Kappa coefficient between those evaluators is 0.62, indicating moderate agreement and confirming the reliability of their assessments.
Overall, \uline{\tech{} outperforms all baselines in terms of Accuracy, Adequacy, and Intention.}
This suggests that \tech{} generates comments that align with human expectations. 
Specifically, by enhancing LLMs' comprehension of questions through knowledge augmentation, \tech{} can generate more accurate comments that better reflect human intent.
However, knowledge augmentation does not necessarily improve the naturalness of the generated answers; thus, FSMIC and \tech{} received similar scores in this aspect.

\begin{table}[t]
\small

\centering
\belowrulesep=0pt
\aboverulesep=0pt
\caption{The results of human evaluation\label{tab:rq4}}
 \begin{adjustbox}{width=1\linewidth,center}
 \begin{threeparttable}


    \begin{tabular}{l|l|cc|l|cc}
    \toprule
    Method & \multicolumn{1}{c|}{Metric} & \multicolumn{1}{c}{Avg.} & \multicolumn{1}{c|}{Std.} & \multicolumn{1}{c|}{Metric} & \multicolumn{1}{c}{Avg.} & \multicolumn{1}{c}{Std.} \\
    \midrule
    DOME  & \multirow{3}{*}{Accuracy} & 3.7   & 1.3   & \multirow{3}{*}{Intention} & 4.1   & 1.1  \\
    FSMIC &       & 4.4   & 0.8   &       & 4.4   & 0.9  \\
    KUMIC &       & \textbf{4.7}   & \textbf{0.7}   &       & \textbf{4.6}   & \textbf{0.6}  \\
    \midrule
    DOME  & \multirow{3}{*}{Adequacy} & 3.3   & 1.3   & \multirow{3}{*}{Naturalness} & 4.4   & 1.0  \\
    FSMIC &       & 4.4   & 0.8   &       & \textbf{4.5}   & 0.8  \\
    KUMIC &       & \textbf{4.5}   & \textbf{0.7}   &       & \textbf{4.5}   & \textbf{0.6}  \\
    \bottomrule
    \end{tabular}%

\end{threeparttable}
\end{adjustbox}
\end{table}

\section{Discussion}

All analyses in this section are conducted using CodeLlama under the 3-shot, semantics-based setting on the TLC dataset.

\subsection{Impact of Hyperparameter}
\begin{figure}[t]
  \centering
  \subfloat[Hyperparameter $k$\label{fig:k}]{
    \includegraphics[width=0.46\linewidth]{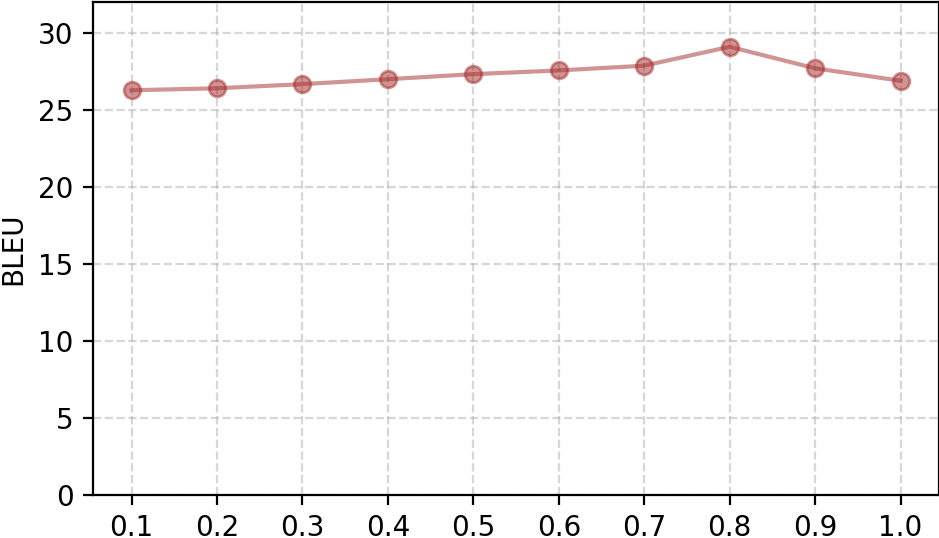}
  }\hfill
  \subfloat[Hyperparameter $p$\label{fig:p}]{
    \includegraphics[width=0.46\linewidth]{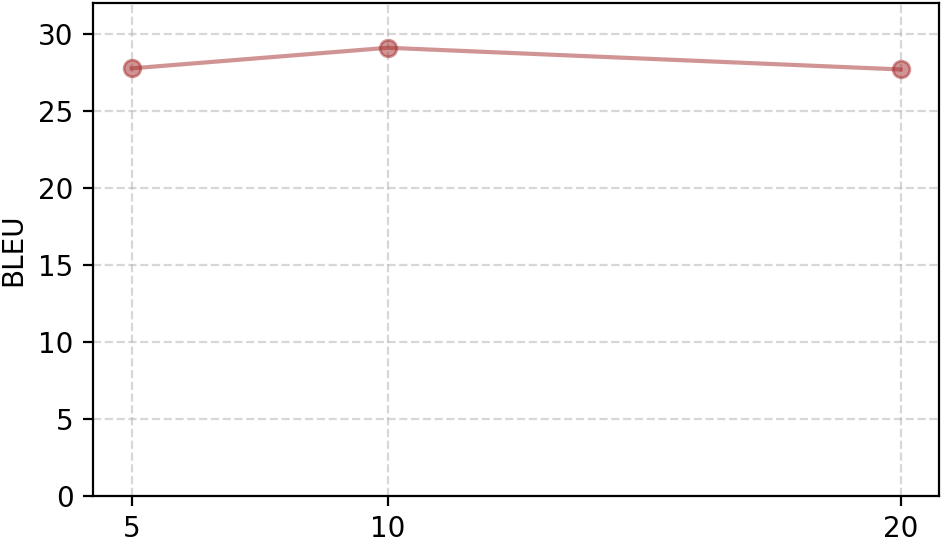}
  }
  \caption{The BLEU values of \tech{} with different hyperparameter values.}
  \label{fig:k_p}
\end{figure}

We investigated the effect of $k$ and $p$ in \tech{} (Fig.\ref{fig:k_p}).
The x-axis denotes the average BLEU score, while the y-axis represents the hyperparameter values.
For $k$, performance steadily increases as the weight of example similarity grows, reaching its highest value within our tested range at $k=0.8$.
This suggests that example similarity contributes more to performance than example quality alone.
For $p$, the best performance within the explored range occurs at $p=10$: when $p=5$, several high-quality examples are excluded, reducing effectiveness, while $p=20$ introduces high-quality but less relevant examples that degrade performance.
Overall, these results indicate that the default configuration of \tech{} performs well within the evaluated range, and that proper tuning of hyperparameters can further enhance its robustness.

\subsection{Impact of Example Quality Assessment}
\begin{figure}[t]
  \centering
  \includegraphics[width=0.58\linewidth]{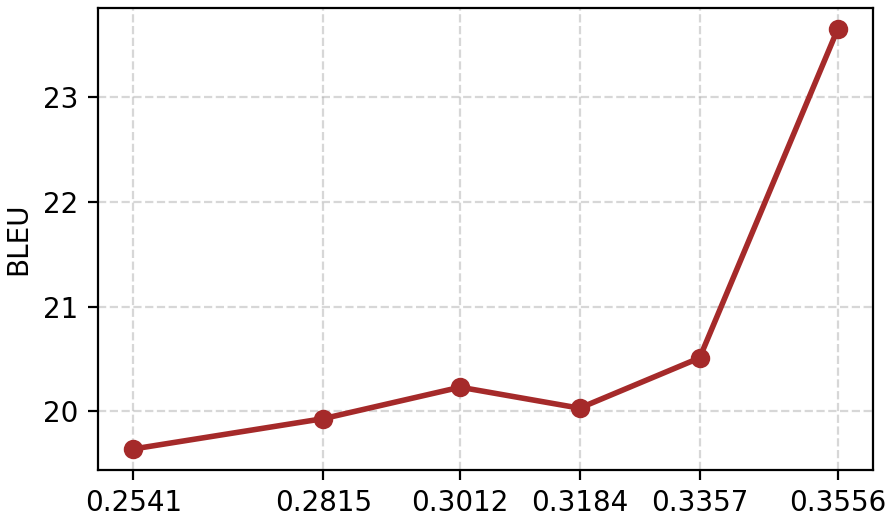}
  \caption{The BLEU values of \tech{} with different quality scores.} 
  \label{fig:dicussion_eq}
\end{figure}

We further analyzed the correlation between example quality and performance of \tech{}.
Specifically, we retrieved the top 18 similar examples, ranked them by quality scores, and partitioned them into six non-overlapping groups of three (Fig.~\ref{fig:dicussion_eq}).
Evaluation under the few-shot setting shows a steady improvement in BLEU scores from 19.64 to 23.65 across groups.
Spearman’s test yields a p-value of 4.8e-3, confirming a significant positive correlation between example quality and \tech{}’s performance.
These results highlight the necessity of prioritizing high-quality examples during retrieval.

\subsection{Reliability of Statements Extraction}
To further assess the reliability of our statement extraction, we  examined whether the extracted statements accurately capture intent-relevant information. 
As current datasets lack ground-truth annotations, we conducted an indirect evaluation by replacing the extracted statements in demonstration examples with (1) the full function code and (2) randomly selected code lines. 
Their average BLEU scores were 25.52 and 25.92, 14.15\% and 12.38\% lower than those of \tech{} in Section~\ref{sec:rq1}
This performance gap indicates that our extraction better captures intent-relevant content and provides more accurate cues for intent-specific comment generation. 
These results provide indirect yet strong evidence of its reliability. 
In future work, we plan to further verify the extraction accuracy through human evaluation against manually annotated statements.

\subsection{Threats To Validity}

\noindent
\textbf{Internal validity}.~
A potential threat is data leakage in LLMs, as LLMs trained on open-source projects may have encountered some test cases. 
However, CodeLlama, Llama3 and Qwen2.5-Coder perform poorly under zero-shot settings, implying their outputs are not purely based on memorization, consistent with prior findings~\cite{DBLP:conf/icse/GengWD00JML24,DBLP:conf/icse/NashidSM23}.
Moreover, inference randomness may affect results; to mitigate this, each experiment was repeated five times and averaged, though residual variability may remain.

\smallskip
\noindent
\textbf{External validity}.~
The primary threat to external validity lies in dataset annotation, as the datasets were annotated by models rather than entirely by human, potentially introducing noise.
To mitigate this, all experiments were conducted on the same datasets to minimize the potential impact of annotation noise.
Another limitation is that our study focuses on Java due to the availability of multi-intent comment datasets.
However, this risk is alleviated by the fact that Java is the most extensively studied language in comment generation, and both datasets are large-scale~\cite{DBLP:conf/icse/GengWD00JML24,DBLP:conf/icse/MuCSWW23,DBLP:conf/sigsoft/RoyFA21,DBLP:conf/kbse/MuC0WW22}. 
Finally, we evaluated only a subset of representative LLMs rather than all available ones, and future work will extend to include larger and more advanced models.

\section{Related Work}

\subsection{Automatic Comment Generation}

Automatic comment generation has long attracted attention in both software engineering and NLP. Early studies relied on templates or information retrieval methods~\cite{DBLP:conf/iwpc/MorenoASMPV13,DBLP:conf/kbse/GrosSDY20,DBLP:conf/kbse/WongYT13,DBLP:conf/wcre/HaiducAMM10}, while recent work adopted neural models trained on large code–text corpora~\cite{DBLP:conf/acl/IyerKCZ16,DBLP:conf/iwpc/HuLXLJ18,DBLP:conf/ijcai/HuLXLLJ18,DBLP:conf/icse/TangSLGHZ022}. 
To overcome data and performance bottlenecks, hybrid approaches combined retrieval and learning, such as retrieving semantically similar code snippets to assist generation~\cite{DBLP:conf/icse/ZhangW00020}.
With the advent of LLMs, several studies enhanced comment generation through prompt augmentation or task-specific guidance~\cite{DBLP:conf/icse/AhmedPDB24v,DBLP:journals/corr/abs-2312-16066}.
For instance, Ahmed et al.\cite{DBLP:conf/icse/AhmedPDB24v} augmented prompts with repository names and detailed function information. 
Moreover, developers often express multiple intents in comments, challenging single-intent comment generation techniques~\cite{DBLP:conf/icse/ZhaiXSTPMXZTZ20, DBLP:journals/tosem/ChenXHLL21, DBLP:conf/icse/MuCSWW23}.
To address this, Mu et al.~\cite{DBLP:conf/icse/MuCSWW23} proposed DOME to model intent-specific patterns, while Geng et al.~\cite{DBLP:conf/icse/GengWD00JML24} explored few-shot prompting with LLMs.
Our work extends this line by explicitly optimizing in-context knowledge and guiding LLMs to focus on intent-related statements for multi-intent comment generation.

\subsection{Automated Code Summary Evaluation}

Automated evaluation metrics for code summarization typically measure similarity between generated and reference summaries.
overlap-based metrics such as BLEU~\cite{DBLP:conf/acl/PapineniRWZ02}, ROUGE~\cite{Lin_2004}, and METEOR~\cite{DBLP:conf/acl/BanerjeeL05} dominate early research, but they fail to capture semantics.
Embedding-based metrics (e.g., BERTScore, SBERT)~\cite{DBLP:conf/emnlp/ReimersG19} improve semantic evaluation, though they remain sensitive to domain variations. 
Recent works further enhances evaluation by incorporating domain-aware or alignment-based criteria~\cite{DBLP:journals/pacmse/JinL24,DBLP:conf/icse/MastropaoloCPB24}.
Jin et al.~\cite{DBLP:journals/pacmse/JinL24} introduced a pre-training task to strengthen domain understanding during model training, while Mastropaolo et al.~\cite{DBLP:conf/icse/MastropaoloCPB24} proposed SIDE, which evaluates the alignment between comments and their corresponding methods.
Inspired by these studies, \tech{} assesses the quality of retrieved examples to ensure that LLMs extract task-relevant knowledge from demonstration examples.

\section{Conclusion}

In this paper, we propose \tech{}, a novel approach for solving the multi-intent comment generation task using LLMs.
Built upon in-context learning, \tech{} leverages Chain-of-Thought (CoT) to optimize knowledge utilization for LLMs to generate intent-specific comments.
Specifically, \tech{} design a retrieval mechanism to retrieve similar demonstration examples that embody intent-related knowledge.
Furthermore, \tech{} employs CoT to guide LLMs in following this knowledge chain: first identifying important statements in the code, and then generating the corresponding intent-specific comments.
Our experiments on two real-world Java datasets demonstrate that \tech{} outperforms state-of-the-art baselines.




\balance
\bibliographystyle{IEEEtran}
\bibliography{ref.bib}

\begin{thebibliography}{10}
\providecommand{\url}[1]{#1}
\csname url@samestyle\endcsname
\providecommand{\newblock}{\relax}
\providecommand{\bibinfo}[2]{#2}
\providecommand{\BIBentrySTDinterwordspacing}{\spaceskip=0pt\relax}
\providecommand{\BIBentryALTinterwordstretchfactor}{4}
\providecommand{\BIBentryALTinterwordspacing}{\spaceskip=\fontdimen2\font plus
\BIBentryALTinterwordstretchfactor\fontdimen3\font minus \fontdimen4\font\relax}
\providecommand{\BIBforeignlanguage}[2]{{%
\expandafter\ifx\csname l@#1\endcsname\relax
\typeout{** WARNING: IEEEtran.bst: No hyphenation pattern has been}%
\typeout{** loaded for the language `#1'. Using the pattern for}%
\typeout{** the default language instead.}%
\else
\language=\csname l@#1\endcsname
\fi
#2}}
\providecommand{\BIBdecl}{\relax}
\BIBdecl

\bibitem{DBLP:journals/tse/XiaBLXHL18}
X.~Xia, L.~Bao, D.~Lo, Z.~Xing, A.~E. Hassan, and S.~Li, ``Measuring program comprehension: {A} large-scale field study with professionals,'' \emph{{IEEE} Trans. Software Eng.}, vol.~44, no.~10, pp. 951--976, 2018.

\bibitem{DBLP:conf/sigdoc/SouzaAO05}
S.~C.~B. de~Souza, N.~Anquetil, and K.~M. de~Oliveira, ``A study of the documentation essential to software maintenance,'' in \emph{{SIGDOC}}.\hskip 1em plus 0.5em minus 0.4em\relax {ACM}, 2005, pp. 68--75.

\bibitem{DBLP:conf/kbse/SridharaHMPV10}
G.~Sridhara, E.~Hill, D.~Muppaneni, L.~L. Pollock, and K.~Vijay{-}Shanker, ``Towards automatically generating summary comments for java methods,'' in \emph{{ASE}}.\hskip 1em plus 0.5em minus 0.4em\relax {ACM}, 2010, pp. 43--52.

\bibitem{DBLP:conf/kbse/LiL000J21}
J.~Li, Y.~Li, G.~Li, X.~Hu, X.~Xia, and Z.~Jin, ``Editsum: {A} retrieve-and-edit framework for source code summarization,'' in \emph{{ASE}}.\hskip 1em plus 0.5em minus 0.4em\relax {IEEE}, 2021, pp. 155--166.

\bibitem{DBLP:journals/tosem/ChenXHLL21}
Q.~Chen, X.~Xia, H.~Hu, D.~Lo, and S.~Li, ``Why my code summarization model does not work: Code comment improvement with category prediction,'' \emph{{ACM} Trans. Softw. Eng. Methodol.}, vol.~30, no.~2, pp. 25:1--25:29, 2021.

\bibitem{DBLP:conf/icse/HuX0WCZ22}
X.~Hu, X.~Xia, D.~Lo, Z.~Wan, Q.~Chen, and T.~Zimmermann, ``Practitioners' expectations on automated code comment generation,'' in \emph{{ICSE}}.\hskip 1em plus 0.5em minus 0.4em\relax {ACM}, 2022, pp. 1693--1705.

\bibitem{DBLP:conf/icse/GengWD00JML24}
M.~Geng, S.~Wang, D.~Dong, H.~Wang, G.~Li, Z.~Jin, X.~Mao, and X.~Liao, ``Large language models are few-shot summarizers: Multi-intent comment generation via in-context learning,'' in \emph{{ICSE}}.\hskip 1em plus 0.5em minus 0.4em\relax {ACM}, 2024, pp. 39:1--39:13.

\bibitem{DBLP:conf/icse/ZhaiXSTPMXZTZ20}
J.~Zhai, X.~Xu, Y.~Shi, G.~Tao, M.~Pan, S.~Ma, L.~Xu, W.~Zhang, L.~Tan, and X.~Zhang, ``{CPC:} automatically classifying and propagating natural language comments via program analysis,'' in \emph{{ICSE}}.\hskip 1em plus 0.5em minus 0.4em\relax {ACM}, 2020, pp. 1359--1371.

\bibitem{DBLP:conf/icse/MuCSWW23}
F.~Mu, X.~Chen, L.~Shi, S.~Wang, and Q.~Wang, ``Developer-intent driven code comment generation,'' in \emph{{ICSE}}.\hskip 1em plus 0.5em minus 0.4em\relax {IEEE}, 2023, pp. 768--780.

\bibitem{DBLP:conf/nips/VaswaniSPUJGKP17}
A.~Vaswani, N.~Shazeer, N.~Parmar, J.~Uszkoreit, L.~Jones, A.~N. Gomez, L.~Kaiser, and I.~Polosukhin, ``Attention is all you need,'' in \emph{{NIPS}}, 2017, pp. 5998--6008.

\bibitem{DBLP:conf/nips/Wei0SBIXCLZ22}
J.~Wei, X.~Wang, D.~Schuurmans, M.~Bosma, B.~Ichter, F.~Xia, E.~H. Chi, Q.~V. Le, and D.~Zhou, ``Chain-of-thought prompting elicits reasoning in large language models,'' in \emph{NeurIPS}, 2022.

\bibitem{DBLP:conf/icse/LeClairJM19}
A.~LeClair, S.~Jiang, and C.~McMillan, ``A neural model for generating natural language summaries of program subroutines,'' in \emph{{ICSE}}.\hskip 1em plus 0.5em minus 0.4em\relax {IEEE} / {ACM}, 2019, pp. 795--806.

\bibitem{DBLP:conf/ijcai/HuLXLLJ18}
X.~Hu, G.~Li, X.~Xia, D.~Lo, S.~Lu, and Z.~Jin, ``Summarizing source code with transferred {API} knowledge,'' in \emph{{IJCAI}}.\hskip 1em plus 0.5em minus 0.4em\relax ijcai.org, 2018, pp. 2269--2275.

\bibitem{our/url}
``Our repository,'' \url{https://github.com/AnonymousWorks312/KUMIC}, 2025.

\bibitem{DBLP:conf/wcre/GolubevPPB21}
Y.~Golubev, V.~Poletansky, N.~Povarov, and T.~Bryksin, ``Multi-threshold token-based code clone detection,'' in \emph{{SANER}}.\hskip 1em plus 0.5em minus 0.4em\relax {IEEE}, 2021, pp. 496--500.

\bibitem{niwattanakul2013using}
S.~Niwattanakul, J.~Singthongchai, E.~Naenudorn, and S.~Wanapu, ``Using of jaccard coefficient for keywords similarity,'' in \emph{Proceedings of the international multiconference of engineers and computer scientists}, vol.~1, no.~6, 2013, pp. 380--384.

\bibitem{DBLP:journals/tosem/ZengYLXWGBDL23}
C.~Zeng, Y.~Yu, S.~Li, X.~Xia, Z.~Wang, M.~Geng, L.~Bai, W.~Dong, and X.~Liao, ``degraphcs: Embedding variable-based flow graph for neural code search,'' \emph{{ACM} Trans. Softw. Eng. Methodol.}, vol.~32, no.~2, pp. 34:1--34:27, 2023.

\bibitem{DBLP:journals/corr/abs-2002-08653}
W.~Wang, G.~Li, B.~Ma, X.~Xia, and Z.~Jin, ``Detecting code clones with graph neural networkand flow-augmented abstract syntax tree,'' \emph{CoRR}, vol. abs/2002.08653, 2020.

\bibitem{DBLP:conf/emnlp/ReimersG19}
N.~Reimers and I.~Gurevych, ``Sentence-bert: Sentence embeddings using siamese bert-networks,'' in \emph{{EMNLP/IJCNLP} {(1)}}.\hskip 1em plus 0.5em minus 0.4em\relax Association for Computational Linguistics, 2019, pp. 3980--3990.

\bibitem{DBLP:journals/pacmse/JinL24}
X.~Jin and Z.~Lin, ``Simllm: Calculating semantic similarity in code summaries using a large language model-based approach,'' \emph{Proc. {ACM} Softw. Eng.}, vol.~1, no. {FSE}, pp. 1376--1399, 2024.

\bibitem{DBLP:conf/icse/MastropaoloCPB24}
A.~Mastropaolo, M.~Ciniselli, M.~D. Penta, and G.~Bavota, ``Evaluating code summarization techniques: {A} new metric and an empirical characterization,'' in \emph{{ICSE}}.\hskip 1em plus 0.5em minus 0.4em\relax {ACM}, 2024, pp. 218:1--218:13.

\bibitem{DBLP:conf/emnlp/WangLGB0H23}
Y.~Wang, H.~Le, A.~Gotmare, N.~D.~Q. Bui, J.~Li, and S.~C.~H. Hoi, ``Codet5+: Open code large language models for code understanding and generation,'' in \emph{{EMNLP}}.\hskip 1em plus 0.5em minus 0.4em\relax Association for Computational Linguistics, 2023, pp. 1069--1088.

\bibitem{DBLP:conf/icse/AhmedPDB24v}
T.~Ahmed, K.~S. Pai, P.~T. Devanbu, and E.~T. Barr, ``Automatic semantic augmentation of language model prompts (for code summarization),'' in \emph{{ICSE}}.\hskip 1em plus 0.5em minus 0.4em\relax {ACM}, 2024, pp. 220:1--220:13.

\bibitem{DBLP:conf/icml/ShrivastavaLT23}
D.~Shrivastava, H.~Larochelle, and D.~Tarlow, ``Repository-level prompt generation for large language models of code,'' in \emph{{ICML}}, ser. Proceedings of Machine Learning Research, vol. 202.\hskip 1em plus 0.5em minus 0.4em\relax {PMLR}, 2023, pp. 31\,693--31\,715.

\bibitem{DBLP:conf/kbse/MuC0WW22}
F.~Mu, X.~Chen, L.~Shi, S.~Wang, and Q.~Wang, ``Automatic comment generation via multi-pass deliberation,'' in \emph{{ASE}}.\hskip 1em plus 0.5em minus 0.4em\relax {ACM}, 2022, pp. 14:1--14:12.

\bibitem{DBLP:conf/kbse/ZhangP0LG22}
J.~Zhang, S.~Panthaplackel, P.~Nie, J.~J. Li, and M.~Gligoric, ``Coditt5: Pretraining for source code and natural language editing,'' in \emph{{ASE}}.\hskip 1em plus 0.5em minus 0.4em\relax {ACM}, 2022, pp. 22:1--22:12.

\bibitem{DBLP:conf/icse/NiuL0GH022}
C.~Niu, C.~Li, V.~Ng, J.~Ge, L.~Huang, and B.~Luo, ``Spt-code: Sequence-to-sequence pre-training for learning source code representations,'' in \emph{{ICSE}}.\hskip 1em plus 0.5em minus 0.4em\relax {ACM}, 2022, pp. 1--13.

\bibitem{DBLP:conf/emnlp/FengGTDFGS0LJZ20}
Z.~Feng, D.~Guo, D.~Tang, N.~Duan, X.~Feng, M.~Gong, L.~Shou, B.~Qin, T.~Liu, D.~Jiang, and M.~Zhou, ``Codebert: {A} pre-trained model for programming and natural languages,'' in \emph{{EMNLP} (Findings)}, ser. Findings of {ACL}, vol. {EMNLP} 2020.\hskip 1em plus 0.5em minus 0.4em\relax Association for Computational Linguistics, 2020, pp. 1536--1547.

\bibitem{DBLP:conf/iclr/GuoRLFT0ZDSFTDC21}
D.~Guo, S.~Ren, S.~Lu, Z.~Feng, D.~Tang, S.~Liu, L.~Zhou, N.~Duan, A.~Svyatkovskiy, S.~Fu, M.~Tufano, S.~K. Deng, C.~B. Clement, D.~Drain, N.~Sundaresan, J.~Yin, D.~Jiang, and M.~Zhou, ``Graphcodebert: Pre-training code representations with data flow,'' in \emph{{ICLR}}.\hskip 1em plus 0.5em minus 0.4em\relax OpenReview.net, 2021.

\bibitem{DBLP:conf/scam/MohammadkhaniTH23}
A.~H. Mohammadkhani, C.~Tantithamthavorn, and H.~Hemmati, ``Explaining transformer-based code models: What do they learn? when they do not work?'' in \emph{{SCAM}}.\hskip 1em plus 0.5em minus 0.4em\relax {IEEE}, 2023, pp. 96--106.

\bibitem{funcom/url}
``Original funcom dataset,'' \url{http://www.ics.uci.edu/lopes/datasets/}, 2010.

\bibitem{DBLP:conf/acl/PapineniRWZ02}
K.~Papineni, S.~Roukos, T.~Ward, and W.~Zhu, ``Bleu: a method for automatic evaluation of machine translation,'' in \emph{{ACL}}.\hskip 1em plus 0.5em minus 0.4em\relax {ACL}, 2002, pp. 311--318.

\bibitem{Lin_2004}
C.-Y. Lin, ``\BIBforeignlanguage{en-US}{Rouge: A package for automatic evaluation of summaries},'' \emph{\BIBforeignlanguage{en-US}{Meeting of the Association for Computational Linguistics,Meeting of the Association for Computational Linguistics}}, Jul 2004.

\bibitem{DBLP:conf/acl/BanerjeeL05}
S.~Banerjee and A.~Lavie, ``{METEOR:} an automatic metric for {MT} evaluation with improved correlation with human judgments,'' in \emph{IEEvaluation@ACL}.\hskip 1em plus 0.5em minus 0.4em\relax Association for Computational Linguistics, 2005, pp. 65--72.

\bibitem{DBLP:conf/iwpc/HaqueEBM22}
S.~Haque, Z.~Eberhart, A.~Bansal, and C.~McMillan, ``Semantic similarity metrics for evaluating source code summarization,'' in \emph{{ICPC}}.\hskip 1em plus 0.5em minus 0.4em\relax {ACM}, 2022, pp. 36--47.

\bibitem{DBLP:journals/corr/abs-2307-09288}
H.~Touvron, L.~Martin, K.~Stone, P.~Albert, A.~Almahairi, Y.~Babaei, N.~Bashlykov, S.~Batra, P.~Bhargava, S.~Bhosale, D.~Bikel, L.~Blecher, C.~Canton{-}Ferrer, M.~Chen, G.~Cucurull, D.~Esiobu, J.~Fernandes, J.~Fu, W.~Fu, B.~Fuller, C.~Gao, V.~Goswami, N.~Goyal, A.~Hartshorn, S.~Hosseini, R.~Hou, H.~Inan, M.~Kardas, V.~Kerkez, M.~Khabsa, I.~Kloumann, A.~Korenev, P.~S. Koura, M.~Lachaux, T.~Lavril, J.~Lee, D.~Liskovich, Y.~Lu, Y.~Mao, X.~Martinet, T.~Mihaylov, P.~Mishra, I.~Molybog, Y.~Nie, A.~Poulton, J.~Reizenstein, R.~Rungta, K.~Saladi, A.~Schelten, R.~Silva, E.~M. Smith, R.~Subramanian, X.~E. Tan, B.~Tang, R.~Taylor, A.~Williams, J.~X. Kuan, P.~Xu, Z.~Yan, I.~Zarov, Y.~Zhang, A.~Fan, M.~Kambadur, S.~Narang, A.~Rodriguez, R.~Stojnic, S.~Edunov, and T.~Scialom, ``Llama 2: Open foundation and fine-tuned chat models,'' \emph{CoRR}, vol. abs/2307.09288, 2023.

\bibitem{llama3/url}
``Introducing meta llama 3: The most capable openly available llm to date,'' \url{https://ai.meta.com/blog/meta-llama-3/}, 2024.

\bibitem{qwen2025qwen25technicalreport}
\BIBentryALTinterwordspacing
Qwen, :, A.~Yang, B.~Yang, B.~Zhang, B.~Hui, B.~Zheng, B.~Yu, C.~Li, D.~Liu, F.~Huang, H.~Wei, H.~Lin, J.~Yang, J.~Tu, J.~Zhang, J.~Yang, J.~Yang, J.~Zhou, J.~Lin, K.~Dang, K.~Lu, K.~Bao, K.~Yang, L.~Yu, M.~Li, M.~Xue, P.~Zhang, Q.~Zhu, R.~Men, R.~Lin, T.~Li, T.~Tang, T.~Xia, X.~Ren, X.~Ren, Y.~Fan, Y.~Su, Y.~Zhang, Y.~Wan, Y.~Liu, Z.~Cui, Z.~Zhang, and Z.~Qiu, ``Qwen2.5 technical report,'' 2025. [Online]. Available: \url{https://arxiv.org/abs/2412.15115}
\BIBentrySTDinterwordspacing

\bibitem{DBLP:conf/icse/SunMLZFLDLC25}
W.~Sun, Y.~Miao, Y.~Li, H.~Zhang, C.~Fang, Y.~Liu, G.~Deng, Y.~Liu, and Z.~Chen, ``Source code summarization in the era of large language models,'' in \emph{{ICSE}}.\hskip 1em plus 0.5em minus 0.4em\relax {IEEE}, 2025, pp. 1882--1894.

\bibitem{DBLP:conf/issta/GuoLWL0HZ023}
H.~Guo, J.~Li, J.~Wang, X.~Liu, D.~Wang, Z.~Hu, R.~Zhang, and H.~Xue, ``Fairrec: Fairness testing for deep recommender systems,'' in \emph{{ISSTA}}.\hskip 1em plus 0.5em minus 0.4em\relax {ACM}, 2023, pp. 310--321.

\bibitem{DBLP:journals/corr/abs-1708-05031}
X.~He, L.~Liao, H.~Zhang, L.~Nie, X.~Hu, and T.~Chua, ``Neural collaborative filtering,'' \emph{CoRR}, vol. abs/1708.05031, 2017.

\bibitem{DBLP:conf/emnlp/ZhouGLZSDLMWD22}
K.~Zhou, Y.~Gong, X.~Liu, W.~X. Zhao, Y.~Shen, A.~Dong, J.~Lu, R.~Majumder, J.~Wen, and N.~Duan, ``Simans: Simple ambiguous negatives sampling for dense text retrieval,'' in \emph{{EMNLP} (Industry Track)}.\hskip 1em plus 0.5em minus 0.4em\relax Association for Computational Linguistics, 2022, pp. 548--559.

\bibitem{DBLP:conf/sigir/FormalLPC22}
T.~Formal, C.~Lassance, B.~Piwowarski, and S.~Clinchant, ``From distillation to hard negative sampling: Making sparse neural {IR} models more effective,'' in \emph{{SIGIR}}.\hskip 1em plus 0.5em minus 0.4em\relax {ACM}, 2022, pp. 2353--2359.

\bibitem{DBLP:conf/uss/0008ZL0C23}
J.~Wang, Z.~Zhang, S.~Liu, X.~Du, and J.~Chen, ``Fuzzjit: Oracle-enhanced fuzzing for javascript engine {JIT} compiler,'' in \emph{{USENIX} Security Symposium}.\hskip 1em plus 0.5em minus 0.4em\relax {USENIX} Association, 2023, pp. 1865--1882.

\bibitem{DBLP:conf/sigsoft/RoyFA21}
D.~Roy, S.~Fakhoury, and V.~Arnaoudova, ``Reassessing automatic evaluation metrics for code summarization tasks,'' in \emph{{ESEC/SIGSOFT} {FSE}}.\hskip 1em plus 0.5em minus 0.4em\relax {ACM}, 2021, pp. 1105--1116.

\bibitem{DBLP:conf/icse/NashidSM23}
N.~Nashid, M.~Sintaha, and A.~Mesbah, ``Retrieval-based prompt selection for code-related few-shot learning,'' in \emph{{ICSE}}.\hskip 1em plus 0.5em minus 0.4em\relax {IEEE}, 2023, pp. 2450--2462.

\bibitem{DBLP:conf/iwpc/MorenoASMPV13}
L.~Moreno, J.~Aponte, G.~Sridhara, A.~Marcus, L.~L. Pollock, and K.~Vijay{-}Shanker, ``Automatic generation of natural language summaries for java classes,'' in \emph{{ICPC}}.\hskip 1em plus 0.5em minus 0.4em\relax {IEEE} Computer Society, 2013, pp. 23--32.

\bibitem{DBLP:conf/kbse/GrosSDY20}
D.~Gros, H.~Sezhiyan, P.~Devanbu, and Z.~Yu, ``Code to comment "translation": Data, metrics, baselining {\&} evaluation,'' in \emph{{ASE}}.\hskip 1em plus 0.5em minus 0.4em\relax {IEEE}, 2020, pp. 746--757.

\bibitem{DBLP:conf/kbse/WongYT13}
E.~Wong, J.~Yang, and L.~Tan, ``Autocomment: Mining question and answer sites for automatic comment generation,'' in \emph{{ASE}}.\hskip 1em plus 0.5em minus 0.4em\relax {IEEE}, 2013, pp. 562--567.

\bibitem{DBLP:conf/wcre/HaiducAMM10}
S.~Haiduc, J.~Aponte, L.~Moreno, and A.~Marcus, ``On the use of automated text summarization techniques for summarizing source code,'' in \emph{{WCRE}}.\hskip 1em plus 0.5em minus 0.4em\relax {IEEE} Computer Society, 2010, pp. 35--44.

\bibitem{DBLP:conf/acl/IyerKCZ16}
S.~Iyer, I.~Konstas, A.~Cheung, and L.~Zettlemoyer, ``Summarizing source code using a neural attention model,'' in \emph{{ACL} {(1)}}.\hskip 1em plus 0.5em minus 0.4em\relax The Association for Computer Linguistics, 2016.

\bibitem{DBLP:conf/iwpc/HuLXLJ18}
X.~Hu, G.~Li, X.~Xia, D.~Lo, and Z.~Jin, ``Deep code comment generation,'' in \emph{{ICPC}}.\hskip 1em plus 0.5em minus 0.4em\relax {ACM}, 2018, pp. 200--210.

\bibitem{DBLP:conf/icse/TangSLGHZ022}
Z.~Tang, X.~Shen, C.~Li, J.~Ge, L.~Huang, Z.~Zhu, and B.~Luo, ``Ast-trans: Code summarization with efficient tree-structured attention,'' in \emph{{ICSE}}.\hskip 1em plus 0.5em minus 0.4em\relax {ACM}, 2022, pp. 150--162.

\bibitem{DBLP:conf/icse/ZhangW00020}
J.~Zhang, X.~Wang, H.~Zhang, H.~Sun, and X.~Liu, ``Retrieval-based neural source code summarization,'' in \emph{{ICSE}}.\hskip 1em plus 0.5em minus 0.4em\relax {ACM}, 2020, pp. 1385--1397.

\bibitem{DBLP:journals/corr/abs-2312-16066}
W.~Sun, C.~Fang, Y.~You, Y.~Chen, Y.~Liu, C.~Wang, J.~Zhang, Q.~Zhang, H.~Qian, W.~Zhao, Y.~Liu, and Z.~Chen, ``A prompt learning framework for source code summarization,'' \emph{CoRR}, vol. abs/2312.16066, 2023.

\end{thebibliography}

\vfill

\end{document}